\documentclass[letter, twocolumn, superscriptaddress, footinbib]{revtex4-1}
\usepackage{amsmath, amssymb, mathtools, etoolbox}
\allowdisplaybreaks[2]
\usepackage{graphicx}
\graphicspath{ {./figures/} }
\usepackage[caption=false,subrefformat=parens,labelformat=parens]{subfig}
\usepackage{sidecap}
\usepackage{ulem} 

\begin{document}

\title{Bet-hedging against Demographic Fluctuations}

\author{BingKan Xue}
\affiliation{The Simons Center for Systems Biology, Institute for Advanced Study, Princeton, NJ 08540.}
\author{Stanislas Leibler}
\affiliation{The Simons Center for Systems Biology, Institute for Advanced Study, Princeton, NJ 08540.}
\affiliation{Laboratory of Living Matter and Center for Studies in Physics and Biology, The Rockefeller University,
New York, NY 10065.}

\date{\today}

\begin{abstract}
Biological organisms have to cope with stochastic variations in both the external environment and the internal population dynamics. Theoretical studies and laboratory experiments suggest that population diversification could be an effective bet-hedging strategy for adaptation to varying environments. Here we show that bet-hedging can also be effective against demographic fluctuations that pose a trade-off between growth and survival for populations even in a constant environment. A species can maximize its overall abundance in the long term by diversifying into coexisting subpopulations of both ``fast-growing'' and ``better-surviving'' individuals. Our model generalizes statistical physics models of birth-death processes to incorporate dispersal, during which new populations are founded, and can further incorporate variations of local environments. In this way we unify different bet-hedging strategies against demographic and environmental variations as a general means of adaptation to both types of uncertainties in population growth.
\end{abstract}

\maketitle

Growth of biological populations is a stochastic process subject to various types of uncertainties. In particular, environmental variations change the growth rate of a population by affecting the physical condition of individual organisms, whereas demographic variations cause the population size to fluctuate due to intrinsic noise in birth and death processes. Such processes have been vigorously studied using statistical physics models \cite{Krapisky2010, Rivoire2011}.

When considering the evolutionary success of a species, it is often assumed that a faster growth rate on average would help a species to achieve greater abundance in the long term. Thus, for example, in a fluctuating environment, a population that has the largest long-term average growth rate is supposed to be the most favored by natural selection. Under some circumstances, the maximum long-term growth rate of a population can be achieved by diversifying into subpopulations of different phenotypes, a mixed strategy known as ``bet-hedging'' \cite{Simons2011, Grimbergen2015}. Many studies have focused on bet-hedging in temporally or spatially varying environments \cite{Kussell2005a, Rajon2009, Carja2014, Patra2014, Hidalgo2015} or under stochastic ecological interactions \cite{Rulands2014}.

Another factor in population dynamics which has been less studied is the extinction risk of local populations. If all individuals in a population happen to die before producing new offspring, the population will go extinct and never recover. The probability that such an extinction event happens can be significant for small populations. This extinction risk is uniquely caused by demographic fluctuations, which exists even in the absence of environmental variations.

The growth rate and the extinction risk of a population may depend differently on the phenotype of individuals. Consider an asexual population whose individuals have a birth rate $\beta$ and a death rate $\delta$. The growth rate of the population is given by $r = \beta - \delta$, while the extinction risk is associated with the factor $q = \delta / \beta$ (it is the probability that a population founded by one individual goes extinct \cite{Krapisky2010}; see \cite{SM1}). Apparently, a large growth rate $r$ does not guarantee a low extinction risk $q$. Intuitively, the growth rate represents the mean of population size change, while the extinction risk is due to fluctuations around that mean. Since it is common to have a trade-off between maximizing the mean and minimizing the fluctuations, one may expect a similar trade-off between growth and survival. Such trade-offs have been studied in many fields including ecology \cite{Tilman1994, Ronce1997, Hanski1999, Metz2001, Crowley2002, Cotto2013}, economics \cite{Markowitz1952}, and engineering \cite{Fleming1975}.

What consequences may this kind of trade-off have on biological populations? One situation where the trade-off between growth and survival will be important is during biological dispersal \cite{Tilman1994, Ronce1997, Hanski1999, Metz2001, Crowley2002, Cotto2013}. Indeed, natural resources are often limited in a local environment, which can only support a finite population size. A successful species would gain abundance by spreading to more locations. During range expansion, new colonies are typically founded by a small number of individuals. In such circumstances, the survival of new populations may be a more important factor than the growth of already established populations. For example, microbes can be dispersed through interactions between their hosts, and may infect new hosts if they successfully establish large growing populations.

Here, we quantitatively analyze the trade-off between growth and survival using a simple statistical physics model. In this model, individuals can grow within local ``patches'', or disperse to new patches. We show that, depending on the dispersal rate, natural selection may favor either a fast growth rate or a low extinction risk of local populations. More importantly, we find that a bet-hedging strategy that generates coexisting subpopulations of fast-growing and better-surviving phenotypes may help a species achieve the maximum abundance in the long term. The emergence of the optimal bet-hedging strategy is solely a consequence of demographic fluctuations. This contrasts with the results of previous studies that considered bet-hedging as a strategy for maximizing long-term growth in varying environments.

\begin{figure*}
\centering
\includegraphics[width=\textwidth]{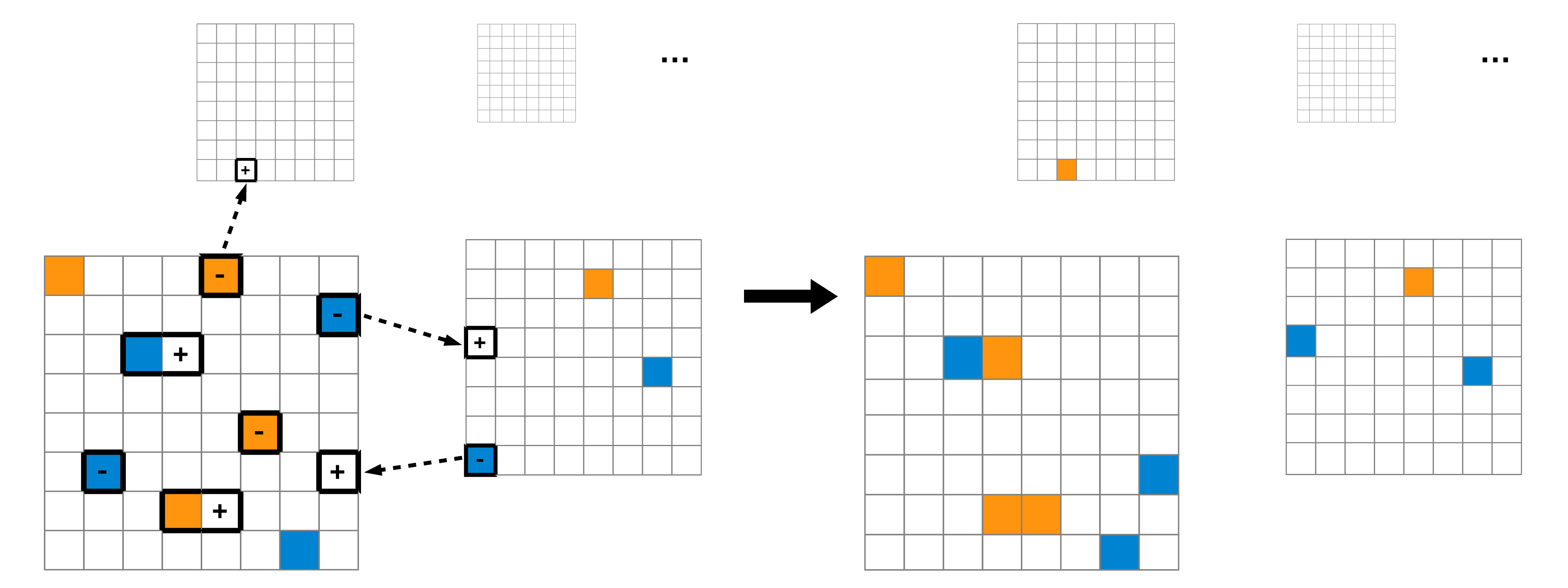}
\caption{\small Schematic illustration of the birth, death, and dispersal processes. Each patch is represented by a grid, and has a finite capacity represented by the number of cells in the grid; an empty cell represents a vacant site, and a colored cell represents an individual, whose phenotype is indicated by its color. The cells highlighted by thick borders are being updated: a cell with a ``$+$'' sign means an individual appears, either being born to another individual (sharing a thick border) or having immigrated from another patch (dashed arrow); a cell with a ``$-$'' sign means an individual disappears, either due to death (isolated thick border) or emigration (dashed arrow). Individuals can move freely within a patch, or disperse to any other patch.} \label{fig:scheme}
\end{figure*}

Our model is illustrated in Fig.~\ref{fig:scheme}. We consider a biological species whose individuals may give birth to a new individual, die, or disperse to another patch. For simplicity, we assume that the patches are equally well connected and have the same carrying capacity $K$. Individuals may express different phenotypes, characterized by different pairs of birth and death rates; dispersal happens passively with a predetermined dispersal rate
\footnote{For models that also consider the dispersal rate as part of a phenotype, see, e.g., \cite{Crowley2002}.}.
For simplicity, consider two phenotypes, $\textrm{A}$ and $\textrm{B}$, which satisfy $r_\textrm{A} \! > \! r_\textrm{B}$ and $q_\textrm{A} \! > \! q_\textrm{B}$, hence phenotype $\textrm{A}$ is fast-growing and phenotype $\textrm{B}$ is better-surviving. To allow for bet-hedging strategies, we assume that each individual randomly expresses one of the phenotypes with probability $\pi_\textrm{A} \! = \! \rho$ and $\pi_\textrm{B} \! = \! 1 \! - \! \rho$ respectively ($0 \! \leq \! \rho \! \leq \! 1$), regardless of its parent's phenotype \cite{SM5}; the phenotype does not change over the lifetime.

\begin{figure}[t]
\centering
\includegraphics[width=0.5\textwidth]{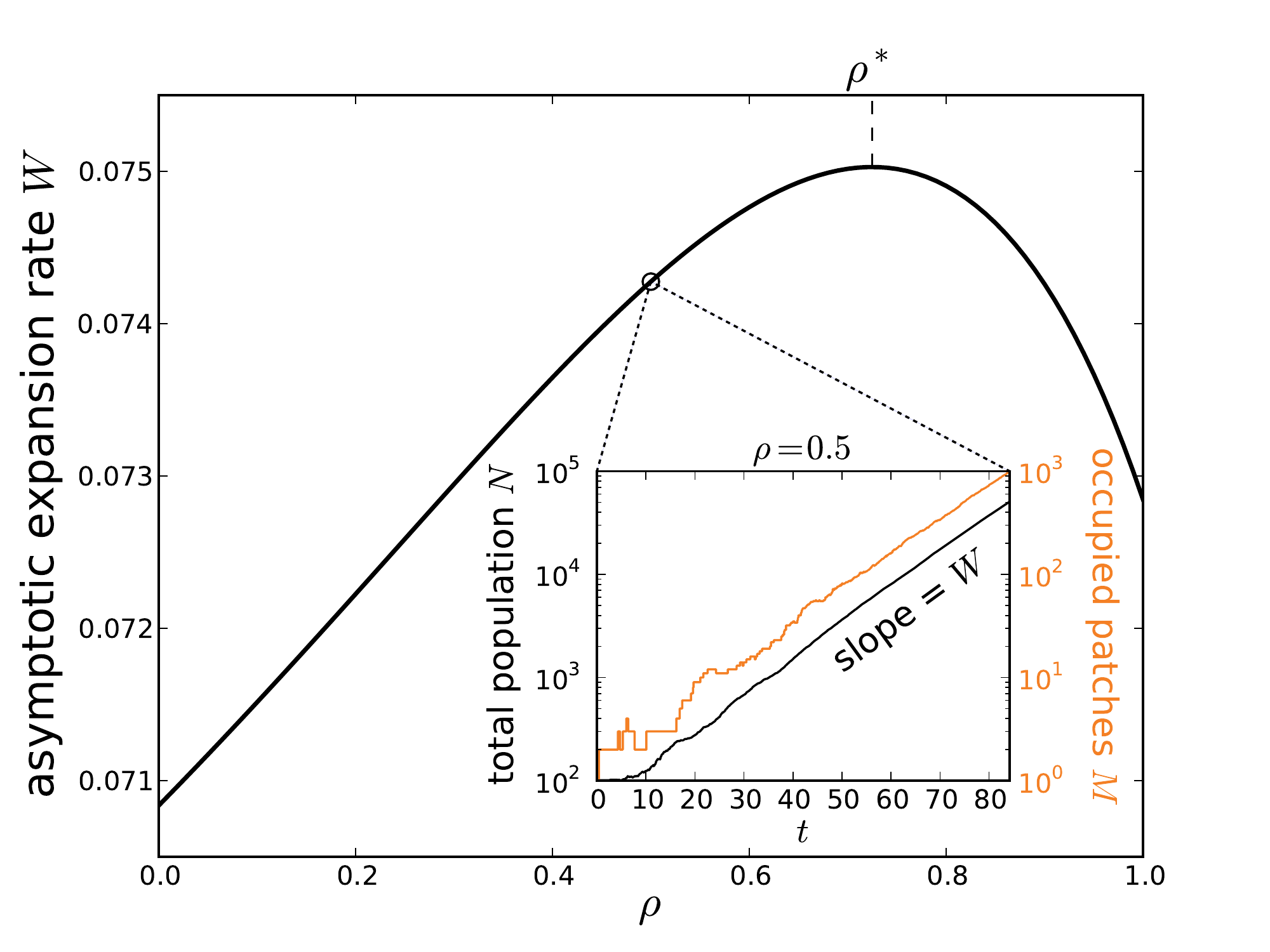}
\caption{\small Asymptotic expansion rate $W$ as a function of phenotype distribution $\rho$. The birth, death, and dispersal rates are $\beta_\textrm{A} = 2$, $\delta_\textrm{A} = 1$, $\beta_\textrm{B} = 0.5$, $\delta_\textrm{B} = 0.1$, $\mu = 0.002$, and the carrying capacity is $K = 100$. Inset: time course of the total population size $N$ and the number of occupied patches $M$, simulated using Gillespie algorithm. The slope of the curves determines the value of $W$ for a given $\rho$.} \label{fig:NinW}
\end{figure}

A typical time course of the total population size is shown in Fig.~\ref{fig:NinW} (inset). The simulation starts with one patch filled with individuals, whose phenotypes are randomly chosen. After an initial phase with relatively large demographic fluctuations, the total population size and the number of occupied patches start to grow at a steady rate. When many patches are available, the species will asymptotically expand at this rate and simultaneously colonize more and more patches. We use this ``asymptotic expansion rate'', $W$, to measure the evolutionary success of the species.

The value of $W$, calculated using \textit{Methods} below, depends on the phenotype distribution $\rho$. Let $\rho^*$ be the value of $\rho$ that maximizes $W$. If $\rho^* \! = \! 1$ or $0$, then a pure strategy with a single phenotype $\textrm{A}$ or $\textrm{B}$ is evolutionarily most successful. However, if the maximum $W$ is reached at an intermediate value $0 \! < \! \rho^* \! < \! 1$, as in Fig.~\ref{fig:NinW}, then a mixed (bet-hedging) strategy, by which a population constantly diversifies into subpopulations of both phenotypes, is more successful in the long term.

\begin{figure}[t]
\centering
\includegraphics[width=0.5\textwidth]{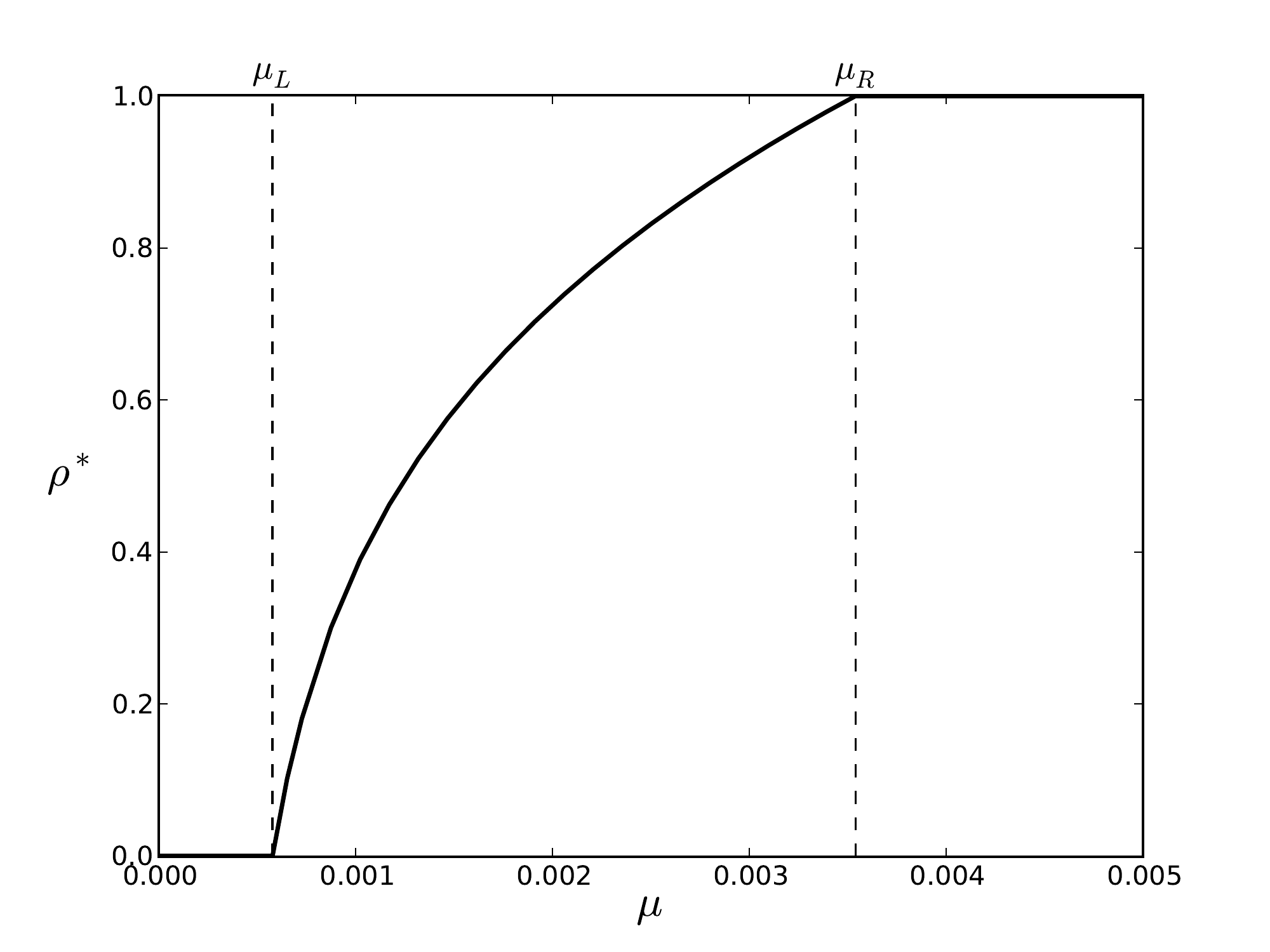}
\caption{Optimal phenotype distribution $\rho^*$ as a function of dispersal rate $\mu$, where one phenotype has a faster growth rate and the other has a lower extinction risk. The birth and death rates of each phenotype are $\beta_\textrm{A} = 2$, $\delta_\textrm{A} = 1$, $\beta_\textrm{B} = 0.5$, $\delta_\textrm{B} = 0.1$; the carrying capacity of each patch is $K = 100$. For dispersal rates between $\mu_L$ and $\mu_R$, a mixed strategy offers the maximum asymptotic expansion rate for the species.}
\label{fig:rho_opt_patchy}
\end{figure}

Those strategies are shown in Fig.~\ref{fig:rho_opt_patchy}, where $\rho^*$ is plotted as a function of the dispersal rate $\mu$. We find $\rho^* = 0$ for $\mu$ below a threshold value, $\mu_L$, and $\rho^* = 1$ above another threshold, $\mu_R$. In these two regimes, a pure strategy of having a single phenotype is thus favored. The fact that each regime favors a different phenotype demonstrates the trade-off between growth and survival. More interestingly, for values of $\mu$ between $\mu_L$ and $\mu_R$, we find a new regime where $0 \! < \! \rho^* \! < \! 1$. In this case, a bet-hedging strategy that produces mixed populations consisting of both phenotypes is evolutionarily favorable. Such a favorable bet-hedging strategy arises only because of intrinsic uncertainties in demographic fluctuations.

So far we have assumed a constant environment for all patches, in contrast to previous studies of bet-hedging that assume a large population in a varying environment
\footnote{But see \cite{Rajon2009, Carja2014} for models that include two local populations and migration between them. See also \cite{Patra2014} for range expansion in a multi-patch environment with nonuniform growth conditions.}.
In the latter case, bet-hedging results from a trade-off between phenotypes that are favorable for different environmental conditions, whereas in our case the trade-off between growth and survival is solely due to demographic fluctuations. These two scenarios can be unified in our model by introducing environmental variations that occur independently for different patches.

For simplicity, assume that there are two possible environmental conditions, $\textrm{X}$ and $\textrm{Y}$, where $\textrm{X}$ is the ``normal'' environment considered above, and $\textrm{Y}$ is a ``hostile'' environment such that the fast-growing phenotype $\textrm{A}$ in environment $\textrm{X}$ becomes unfavorable in $\textrm{Y}$, while the better-surviving phenotype $\textrm{B}$ is unaffected. A good example is bacterial populations that produce both normal cells which thrive in growth media but die under antibiotic treatment and persister cells which are slow-growing but tolerant to antibiotics \cite{Balaban2004}. Thus, the birth and death rates of the two phenotypes satisfy $\beta_\textrm{A}^{(\textrm{X})} \! > \! \beta_\textrm{A}^{(\textrm{Y})} \! = \! 0$, $\delta_\textrm{A}^{(\textrm{Y})} \! \gg \! \delta_\textrm{A}^{(\textrm{X})}$, $\beta_\textrm{B}^{(\textrm{Y})} \! = \! \beta_\textrm{B}^{(\textrm{X})}$, and $\delta_\textrm{B}^{(\textrm{Y})} \! = \! \delta_\textrm{B}^{(\textrm{X})}$. Each patch switches randomly between the two environmental conditions, with switching rates $\alpha_\textrm{X}$ ($\textrm{Y} \to \textrm{X}$) and $\alpha_\textrm{Y}$ ($\textrm{X} \to \textrm{Y}$). Hence, the stationary distribution of the environment is $p_\textrm{X} \! = \! \alpha_\textrm{X} / (\alpha_\textrm{X} \! + \! \alpha_\textrm{Y}) \equiv \epsilon$ and $p_\textrm{Y} \! = \! 1 \! - \! \epsilon$. We assume that the carrying capacity $K$ and the dispersal rate $\mu$ do not depend on the local environment.

\begin{figure}
\centering
\includegraphics[width=0.5\textwidth]{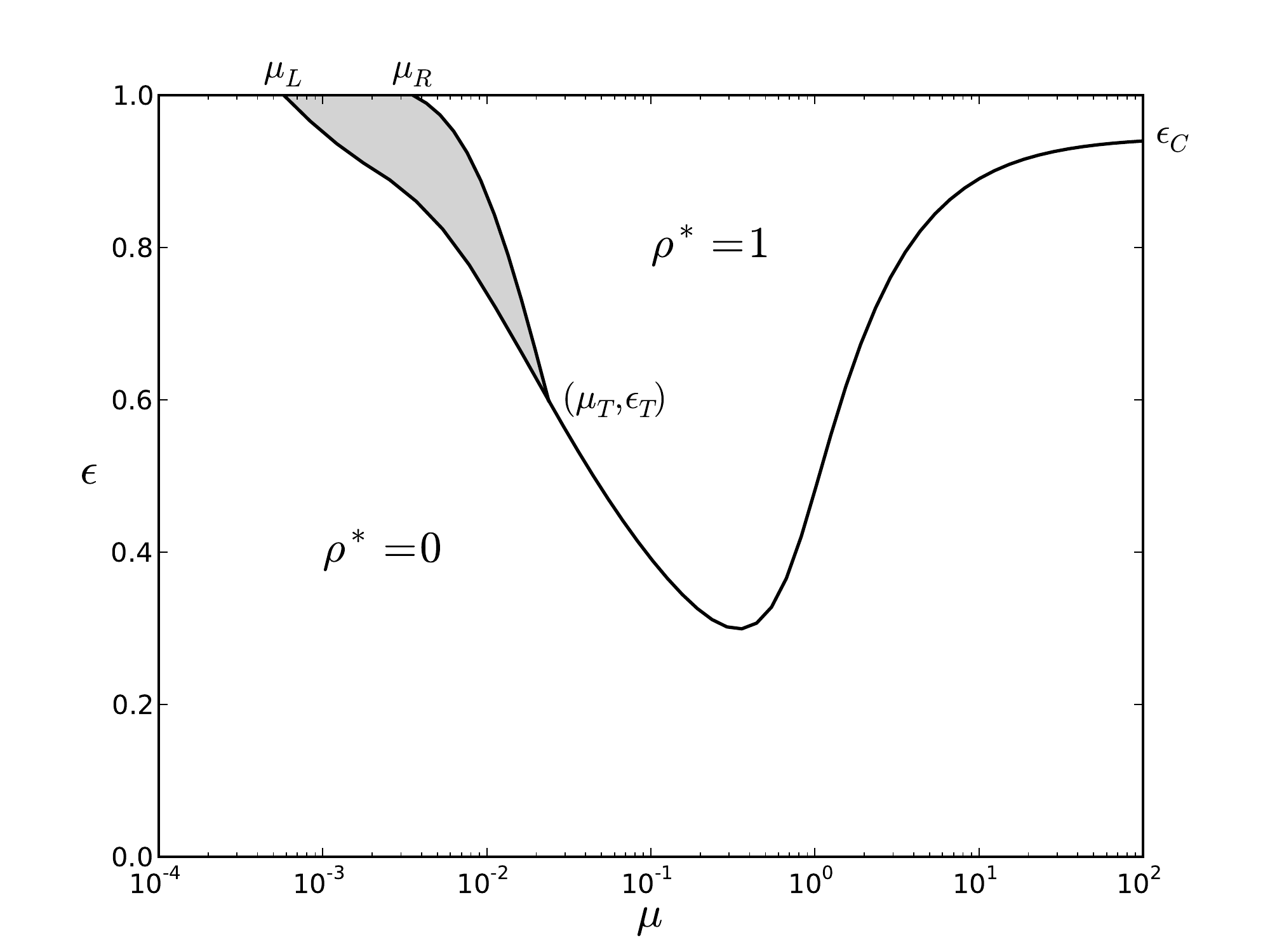}
\caption{Optimal phenotype distribution $\rho^*$ for different values of the dispersal rate $\mu$ and the environment distribution $\epsilon$. Shaded region marks a $0 \! < \! \rho^* \! < \! 1$ phase in which a mixed strategy offers the maximum asymptotic expansion rate. The birth and death rates are $\beta_\textrm{A}^{(\textrm{X})} \! = \! 2$, $\delta_\textrm{A}^{(\textrm{X})} \! = \! 1$, $\beta_\textrm{A}^{(\textrm{Y})} \! = \! 0$, $\delta_\textrm{A}^{(\textrm{Y})} \! = \! 10$, $\beta_\textrm{B}^{(\textrm{X})} \! = \! \beta_\textrm{B}^{(\textrm{Y})} \! = \! 0.5$, $\delta_\textrm{B}^{(\textrm{X})} \! = \! \delta_\textrm{B}^{(\textrm{Y})} \! = \! 0.1$. Each patch has carrying capacity $K \! = \! 100$ and environment switching rates $\alpha \! = \! 0.1$.}
\label{fig:phase}
\end{figure}

To see the effect of local environmental variations, we vary the environment distribution $\epsilon$ while keeping $\alpha \! \equiv \! \alpha_\textrm{X} \! + \! \alpha_\textrm{Y}$ fixed. The optimal phenotype distribution $\rho^*$ that maximizes $W$ now depends on both $\mu$ and $\epsilon$. This can be characterized by a ``phase diagram'' shown in Fig.~\ref{fig:phase} \cite{SM3}. Again, there are three regimes corresponding to pure strategies $\rho^* \! = \! 0$ or $1$, and a mixed strategy $0 \! < \! \rho^* \! < \! 1$. The top edge ($\epsilon \! = \! 1$) corresponds to the case where the environment is $\textrm{X}$ at all times, the same as in Fig.~\ref{fig:rho_opt_patchy}, with a $0 \! < \! \rho^* \! < \! 1$ phase in between the threshold values $\mu_L$ and $\mu_R$. This mixed phase extends to smaller values of $\epsilon$, until reaching a point $(\mu_T, \epsilon_T)$, where it disappears and is replaced by a sharp boundary between the pure phases $\rho^* \! = \! 0$ and $1$.

The topology of the phase diagram can be largely inferred from the behavior of $\rho^*$ near the edges of the diagram, i.e., in the limits $\epsilon \to 0$ or $1$ and $\mu \to 0$ or $\infty$, which can be found analytically \cite{SM2}. Intuitively, near the bottom edge ($\epsilon \! \to \! 0$), $\rho^* \! = \! 0$ since phenotype $\textrm{A}$ is unfit for a constantly hostile environment $\textrm{Y}$. On the far left ($\mu \! < \! \mu_L$), $\rho^* \! = \! 0$ for all $\epsilon$ because phenotype $\textrm{A}$ is disadvantageous even if the environment is always favorable ($\epsilon \! = \! 1$). On the far right ($\mu \! \gg \! \mu_R$), there is a threshold value $\epsilon_C$, above which $\rho^* \! = \! 1$ and below which $\rho^* \! = \! 0$ \cite{SM2}. By smoothly interpolating the phases of $\rho^*$ from the edges to the middle of the diagram, we recover the shape of Fig.~\ref{fig:phase}.

The phase diagram shows that different bet-hedging strategies against demographic and environmental variations are special cases of a general adaptation strategy against both types of uncertainties. The shape of the diagram changes depending on the phenotypes and the environments \cite{SM4}. In extreme cases, bet-hedging could arise mainly because of demographic fluctuations, such as when phenotype $\textrm{A}$ is fast-growing and phenotype $\textrm{B}$ is better-surviving in both environments $\textrm{X}$ and $\textrm{Y}$ (Fig.~S3a). Alternatively, bet-hedging may be optimal mainly due to environmental variations, such as when phenotype $\textrm{A}$ is both fast-growing and better-surviving in environment $\textrm{X}$, whereas phenotype $\textrm{B}$ is fast-growing and better-surviving in $\textrm{Y}$ (Fig.~S3b).

Our results indicate the generality of bet-hedging as a means of coping with various types of uncertainties encountered by biological populations. In reality, organisms live in much more complex environments and interact with many other species --- their habitats may be spatially structured, and their dispersal may be affected by ecological conditions \cite{Hanski1999, Ronce1997, Metz2001, Crowley2002, Cotto2013}. Nevertheless, our simple model, which shows that organisms can bet-hedge even against purely stochastic demographic fluctuations, clearly suggests a broader perspective for understanding the advantage of bet-hedging behavior widely observed in nature.

The trade-off between growth and survival exists in many situations and is not particular to the simple dispersal process considered in our model. The idea that bet-hedging can be effective against stochastic fluctuations due to small numbers is applicable to other fields, such as financial investment and information processing.

\medskip
\textit{Methods.}---
Here we briefly describe methods for calculating the asymptotic expansion rate $W$ and the optimal phenotype distribution $\rho^*$ for the basic model with constant environment; technical details and generalization to locally fluctuating environments are given in \cite{SM2}.

For a given dispersal rate $\mu$, $W$ is calculated by considering the following ``patch dynamics'' \cite{SM2}. Let a patch be labeled by the type $\textrm{P}_{nl}$ if it has $n$ individuals, $l$ of which have phenotype $\textrm{A}$ and $n \! - \! l$ phenotype $\textrm{B}$, satisfying $0 \! \leq \! n \! \leq \! K$ and $0 \! \leq \! l \! \leq \! n$. Let $m_{nl}$ be the number of such patches, then the number of occupied patches is $M \! = \! \sum_{n,l} m_{nl}$, and the total population size is $N \! = \! \sum_{n,l} n \, m_{nl}$. The birth, death, and dispersal processes can be described in terms of the patches as:
\begin{align}
\textrm{P}_{n,l} &\xrightarrow{\mathmakebox[0.5in]{\rho \, \beta_{nl}}} \textrm{P}_{n+1,l+1} \; , \label{eq:patch-birthA} \\
\textrm{P}_{n,l} &\xrightarrow{\mathmakebox[0.5in]{(1-\rho) \beta_{nl}}} \textrm{P}_{n+1,l} \; , \label{eq:patch-birthB} \\
\textrm{P}_{n,l} &\xrightarrow{\mathmakebox[0.5in]{\gamma_{nl}}} \textrm{P}_{n-1,l-1} \; , \label{eq:patch-deathA} \\
\textrm{P}_{n,l} &\xrightarrow{\mathmakebox[0.5in]{\delta_{nl}}} \textrm{P}_{n-1,l} \; , \label{eq:patch-deathB} \\
\textrm{P}_{n,l} &\xrightarrow{\mathmakebox[0.5in]{\mu_{l}}} \textrm{P}_{n-1,l-1} + \textrm{P}_{1,1} \; , \label{eq:patch-migrationA} \\
\textrm{P}_{n,l} &\xrightarrow{\mathmakebox[0.5in]{\mu_{n-l}}} \textrm{P}_{n-1,l} + \textrm{P}_{1,0} \; , \label{eq:patch-migrationB}
\end{align}
where the rate constants are $\beta_{nl} = \big( \beta_\textrm{A} l + \beta_\textrm{B} (n-l) \big) \, (1 \! - \! n/K)$, $\gamma_{nl} = \delta_\textrm{A} \, l \, (1 \! - \! n/K)$, $\delta_{nl} = \delta_\textrm{B} \, (n \! - \! l) \, (1 \! - \! n/K)$, and $\mu_n = \mu \, n$. For (\ref{eq:patch-migrationA}, \ref{eq:patch-migrationB}), we have assumed that there is a large supply of available patches, so that a dispersed individual would always end up in an empty patch. The deterministic dynamics of the patch numbers is given by
\begin{align} \label{eq:patch-mixed}
&\dot{m}_{n,l} = \rho \, \beta_{n-1,l-1} \, m_{n-1,l-1} + (1-\rho) \, \beta_{n-1,l} \, m_{n-1,l} \nonumber \\
&+ (\gamma_{n+1,l+1} \! + \! \mu_{l+1}) m_{n+1,l+1} + (\delta_{n+1,l} \! + \! \mu_{n-l+1}) m_{n+1,l} \nonumber \\
&- (\beta_{n,l} + \gamma_{n,l} + \delta_{n,l} + \mu_n) m_{n,l} \nonumber \\
&+ \delta_{n,1} \, {\textstyle \sum_{n'=1}^{K} \sum_{l'=0}^{n'}} (\delta_{l,1} \, \mu_{l'} + \delta_{l,0} \, \mu_{n'-l'}) \, m_{n',l'} \; .
\end{align}
This equation can be cast in a matrix form (with composite indices), $\dot{m}_{nl} = \sum_{n'l'} H_{nl,n'l'} \, m_{n'l'}$. The largest real eigenvalue of the matrix $\boldsymbol{H}$ yields $W$, which can then be numerically maximized over $\rho$ to find $\rho^*$ \cite{SM3}.

Approximate expressions of $W$ can be obtained in the limit where the dispersal rate $\mu$ is large or small \cite{SM2}. For $\mu \gg \beta_a$ and $\delta_a$, where $a \! = \! \textrm{A}$ or $B$, we find $W \approx r_\textrm{m}$, where $r_\textrm{m}(\rho)$ is the growth rate of a mixed population with phenotype distribution $\rho$. Intuitively, when the dispersal rate is high, individuals move freely between the patches, so the whole species behaves as one large population, and hence $W$ is given by $r_\textrm{m}$. Now, $r_\textrm{m}(\rho)$ increases with $\rho$, since the higher the percentage of phenotype $\textrm{A}$ is, the faster the population grows. Therefore, in this regime, maximizing $W$ yields $\rho^* \! = \! 1$, which means the fast-growing phenotype $\textrm{A}$ is favored.

On the other hand, for $\mu \! \ll \! \beta_a/K$ and $\delta_a/K$, we find $W \approx \mu K (1 \! - \! q_\textrm{m})$, where $q_\textrm{m}(\rho)$ is the probability that a mixed population founded by one individual goes extinct before reaching the carrying capacity \cite{SM2}. Intuitively, when the dispersal rate is low, the occupied patches are mostly full, hence the overall rate of dispersal to new patches is proportional to $\mu K$; among those colonization attempts, only a fraction escapes local extinction, hence the factor $(1 \! - \! q_\textrm{m})$. But $q_\textrm{m}(\rho)$ increases with $\rho$ because the larger the proportion of phenotype $\textrm{B}$, the lower the extinction risk. Hence, in this regime, $W$ is maximized by $\rho^* \! = \! 0$, favoring the better-surviving phenotype $\textrm{B}$.

The value of $\rho^*$ increases continuously from $0$ to $1$ as $\mu$ varies between two thresholds $\mu_L$ and $\mu_R$. The values of $\mu_L$ and $\mu_R$ depend on the parameters $(\beta_a, \delta_a)$. For a wide range of parameters that satisfy $r_\textrm{A} \! > \! r_\textrm{B}$ and $q_\textrm{A} \! > \! q_\textrm{B}$, the $0 \! < \! \rho^* \! < \! 1$ regime exists \cite{SM3}, where a bet-hedging strategy is most favorable.

\medskip \noindent
\textbf{Acknowledgments.}
We thank Edo Kussell and Luca Peliti for extremely helpful discussions and comments. This research has been partly supported by grants from the Simons Foundation to S.L. through the Rockefeller University (Grant No. 345430) and the Institute for Advanced Study (Grant No. 345801). B.X. is funded by the Eric and Wendy Schmidt Membership in Biology at the Institute for Advanced Study.

\AtEndEnvironment{thebibliography}{
\bibitem{SM1}%
  See Supplemental Material Sec.~I~A for an elementary derivation of the extinction probability $q$ in the birth and death processes \cite{Lohmar2011, Houchmandzadeh2010a}; Sec.~I~B generalizes the results to the case of a mixed strategy with two phenotypes.

\bibitem{SM2}%
  See Supplemental Material Sec.~II~A for details of deriving the patch dynamics and calculating the asymptotic expansion rate $W$ (similar methods are used in treating structured metapopulation models \cite{Hanski1999, Metz2001}); Sec.~II~B generalizes the method to the case of mixed populations; Sec.~II~C considers the case of patchy and fluctuating environments.

\bibitem{SM3}%
  See Supplemental Material Sec.~III for numerical calculations of the optimal phenotype distribution $\rho^*$.

\bibitem{SM4}%
  See Supplemental Material Sec.~IV for more examples of the possible phase diagrams.

\bibitem{SM5}%
  See Supplemental Material Sec.~V for a generalization of the model where the distribution of individual phenotypes may depend on the parental phenotype.
}

\bibliography{new2}

\clearpage

\onecolumngrid
\begin{center}
{\large \bf
Supplemental Material for\\
``Bet-hedging against Demographic Fluctuations''
}\\
\bigskip
BingKan Xue$^1$ and Stanislas Leibler$^{1,2}$\\
\medskip
{\it $^1$ The Simons Center for Systems Biology, Institute for Advanced Study, Princeton, NJ 08540.\\
$^2$ Laboratory of Living Matter and Center for Studies in Physics and Biology,\\
The Rockefeller University, New York, NY 10065.}
\end{center}
\vspace{15pt}
\twocolumngrid

\renewcommand{\theequation}{S\arabic{equation}}
\renewcommand{\thefigure}{S\arabic{figure}}
\setcounter{equation}{0}
\setcounter{figure}{0}

\section{Extinction probability $Q$} \label{app:extinction}

Here we derive the probability that a local population founded by a small number of individuals goes extinct before reaching the carrying capacity. We first treat the case of a single phenotype, and then generalize to the case of a mixed strategy with two phenotypes.

\subsection{single phenotype} \label{app:ex-single}

For a single phenotype, the birth and death processes can be described by
\begin{align}
\mathbb{I} + \emptyset &\xrightarrow{\beta/K} 2 \, \mathbb{I} , \label{eq:b} \\
\mathbb{I} + \emptyset &\xrightarrow{\delta/K} 2 \, \emptyset , \label{eq:d}
\end{align}
where $\mathbb{I}$ represents an individual and $\emptyset$ represents a vacancy in a local patch with carrying capacity $K$. The rate constants are scaled such that, under mass-action kinetics, the birth and death rates per capita are $\beta$ and $\delta$ when the population size is small ($n \ll K$). Eq.~(\ref{eq:b}) means an individual has to find a vacancy within the patch in order to reproduce, which ensures that the local population size does not exceed the carrying capacity. Eq.~(\ref{eq:d}) is analogous to a continuous time Moran process --- an individual dies by being replaced by a vacancy, which ensures that a fully occupied patch is stable.

This last assumption simplifies the analysis while keeping the main features of the birth-death processes. Indeed, compare that to an alternative death process described by $\mathbb{I} \xrightarrow{\delta} \emptyset$, where the death rate per capita does not depend on the population size. The latter process, together with Eq.~(\ref{eq:b}), would lead to a deterministic dynamics of the population size given by $\dot{n} = \beta \, n (1-n/K) - \delta \, n$. This is equivalent to $\dot{n} = r \, n (1 - n/\tilde{K})$, where $r = \beta - \delta$ is the growth rate, and $\tilde{K} = K r/\delta$ is the effective carrying capacity. Therefore, at the level of deterministic dynamics, the latter death process simply amounts to redefining the carrying capacity $K$. On the stochastic level, however, the latter process allows the population size to fluctuate even after reaching the carrying capacity, which results in a nonzero rate of extinction due to large fluctuations \cite{Lohmar2011}. Nevertheless, this rate of extinction is exponentially small in $K$, which is well beyond the timescale of local extinction during range expansion considered in this paper (see below). Therefore, neglecting such instability of fully occupied patches allows us to use Eq.~(\ref{eq:d}) without affecting the results.

Let $\mathbb{P}_n(t)$ be the probability that the population size is $n$ at time $t$, where $0 \leq n \leq K$. Then $\mathbb{P}_n(t)$ obeys the master equation:
\begin{align}
\frac{d \mathbb{P}_n}{dt} &= \beta \, (n-1) \, \big( 1 - (n-1)/K \big) \, \mathbb{P}_{n-1} \nonumber \\
&\quad + \delta \, (n+1) \, \big( 1 - (n+1)/K \big) \, \mathbb{P}_{n+1} \nonumber \\
&\quad - (\beta + \delta) \, n \, \big( 1 - n/K \big) \, \mathbb{P}_{n} .
\end{align}
It can be solved by using the generating function \cite{Houchmandzadeh2010a}
\begin{equation}
\Phi(z,t) \equiv \sum_{n=0}^{K} \mathbb{P}_n(t) \, z^n ,
\end{equation}
which obeys the equation
\begin{equation} \label{eq:phi-single}
\frac{\partial \Phi}{\partial t} = (1-z) (\delta - \beta z) \, \frac{\partial}{\partial z} \Big( \Phi - \frac{z}{K} \, \frac{\partial \Phi}{\partial z} \Big) .
\end{equation}
It can be seen that $\frac{\partial \Phi}{\partial t} = 0$ for $z = 1$ and $z = q \equiv \delta/\beta$ (we assume $0 < \delta < \beta$, so $0 < q < 1$), hence, for all $t \geq 0$,
\begin{align}
\Phi(1,t) &= \textrm{const} = 1 , \label{eq:bc1} \\
\Phi(q,t) &= \textrm{const} = \Phi(q,0) . \label{eq:bc2}
\end{align}

Suppose initially there are $n_0$ individuals ($n_0 \ll K$), so that $\Phi(z,0)$ is given by
\begin{equation}
\Phi(z,0) = z^{n_0} .
\end{equation}
For a finite $K$, the population size will eventually reach a stationary distribution, and $\Phi(z,t)$ converges to a function $\Phi_\textrm{s}(z)$ satisfying
\begin{equation}
\frac{d}{dz} \Big( \Phi_\textrm{s} - \frac{z}{K} \, \frac{d \Phi_\textrm{s}}{d z} \Big) = 0 .
\end{equation}
The solution given by boundary conditions (\ref{eq:bc1}, \ref{eq:bc2}) is
\begin{equation}
\Phi_\textrm{s}(z) = \frac{q^{n_0} - q^K}{1 - q^K} + \frac{1 - q^{n_0}}{1 - q^K} \, z^K .
\end{equation}

Finally, the probability of extinction is given by
\begin{align}
Q &= \mathbb{P}_0(t\to\infty) = \Phi(0,t\to\infty) = \Phi_\textrm{s}(0) \nonumber \\
&= \frac{q^{n_0} - q^K}{1 - q^K} \approx q^{n_0} , \quad \textrm{for } K \gg n_0 \geq 1.
\end{align}
In particular, $q = \delta/\beta$ is the extinction risk of a local population founded by one individual. Note that $\Phi(z,t)$ converges to $\Phi_\textrm{s}(z)$ almost exponentially fast over a timescale $\sim 1/(\beta-\delta)$, hence extinction happens most likely within a few generations' time, long before the population approaches the carrying capacity.

\subsection{mixed population} \label{app:ex-mixed}

Now consider the case of a mixed population with two phenotypes $\{ \textrm{A}, \textrm{B} \}$, and a constant phenotype distribution $\rho$. The birth and death processes are described by
\begin{align}
\mathbb{I}^a + \emptyset &\xrightarrow{\mathmakebox[0.5in]{\pi_b \, \beta_a /K}} \mathbb{I}^a + \mathbb{I}^b , \label{eq:birth-mixed} \\
\mathbb{I}^a + \emptyset &\xrightarrow{\mathmakebox[0.5in]{\delta_a /K}} 2 \, \emptyset , \label{eq:death-mixed}
\end{align}
where $\beta_a$ and $\delta_a$ are the birth and death rates for each phenotype $\mathbb{I}^a$, where $a = \textrm{A}, \textrm{B}$, and $\pi_\textrm{A} = \rho$, $\pi_\textrm{B} = 1-\rho$.

Let $\mathbb{P}_{m,n}(t)$ be the probability that there are $m$ individuals with phenotype $\textrm{A}$ and $n$ with phenotype $\textrm{B}$ at time $t$, where $0 \leq m, n \leq K$ and $0 \leq m+n \leq K$. Similarly to the single phenotype case, we define a generating function
\begin{equation}
\Phi(x,y,t) \equiv \sum_{m=0}^{K} \sum_{n=0}^{K-m} \mathbb{P}_{m,n}(t) \, x^m \, y^n ,
\end{equation}
which obeys the equation
\begin{align} \label{eq:phi-mixed}
\frac{\partial \Phi}{\partial t} = \; &\bigg[ \Big( (1-x) \big( \delta_\textrm{A} - \rho \beta_\textrm{A} x \big) - (1-y) (1-\rho) \beta_\textrm{A} x \Big) \frac{\partial}{\partial x} \nonumber \\
&+ \Big( (1-y) \big( \delta_\textrm{B} - (1-\rho) \beta_\textrm{B} y \big) - (1-x) \rho \beta_\textrm{B} y \Big) \frac{\partial}{\partial y} \bigg] \nonumber \\
&\Big( \Phi - \frac{x}{K} \, \frac{\partial \Phi}{\partial x} - \frac{y}{K} \, \frac{\partial \Phi}{\partial y} \Big) .
\end{align}
There are three fixed points where $\frac{\partial \Phi}{\partial t} = 0$, given by the set of equations
\begin{equation}
\hspace{-2pt} \left\{ \begin{array}{l}
(1-x) \big( \delta_\textrm{A} - \rho \beta_\textrm{A} x \big) - (1-y) (1-\rho) \beta_\textrm{A} x = 0 , \\[4pt]
(1-x) \rho \beta_\textrm{B} y - (1-y) \big( \delta_\textrm{B} - (1-\rho) \beta_\textrm{B} y \big) = 0 .
\end{array} \right.
\end{equation}
Besides the trivial solution $(x,y) = (1,1)$ which implies normalization, $\Phi(1,1,t) = 1$, the other two are
\begin{align}
& \left\{ \begin{array}{l}
x_\textrm{s} = \frac{2 q_\textrm{A}}{(1 + q_\textrm{A} - q_\textrm{B}) + \sqrt{(1 - q_\textrm{A} + q_\textrm{B})^2 + 4 (1-\rho) (q_\textrm{A} - q_\textrm{B})}} \, , \\[4pt]
y_\textrm{s} = \frac{2 q_\textrm{B}}{(1 - q_\textrm{A} + q_\textrm{B}) + \sqrt{(1 - q_\textrm{A} + q_\textrm{B})^2 + 4 (1-\rho) (q_\textrm{A} - q_\textrm{B})}} \, ;
\end{array} \right. \label{eq:xsys} \\[6pt]
& \left\{ \begin{array}{l}
x_\textrm{l} = \frac{(1 + q_\textrm{A} - q_\textrm{B}) + \sqrt{(1 - q_\textrm{A} + q_\textrm{B})^2 + 4 (1-\rho) (q_\textrm{A} - q_\textrm{B})}}{2 \rho (q_\textrm{A} - q_\textrm{B}) / q_\textrm{A}} \, , \\[4pt]
y_\textrm{l} = \frac{(1 - q_\textrm{A} + q_\textrm{B}) + \sqrt{(1 - q_\textrm{A} + q_\textrm{B})^2 + 4 (1-\rho) (q_\textrm{A} - q_\textrm{B})}}{2 (1-\rho) (q_\textrm{A} - q_\textrm{B}) / q_\textrm{B}} \, ,
\end{array} \right.
\end{align}
where $q_\textrm{A} \equiv \delta_\textrm{A} / \beta_\textrm{A}$ and $q_\textrm{B} \equiv \delta_\textrm{B} / \beta_\textrm{B}$. It follows that
\begin{align} \label{eq:bc-s}
\Phi(x_\textrm{s}, y_\textrm{s}, t) = \Phi(x_\textrm{s}, y_\textrm{s}, 0) = x_\textrm{s}^{m_0} \, y_\textrm{s}^{n_0} ,
\end{align}
and similarly for $(x_\textrm{l}, y_\textrm{l})$, where $m_0, n_0$ are the initial numbers of individuals with phenotype $\textrm{A}$ and $\textrm{B}$ respectively. Note that $0 < x_\textrm{s}, y_\textrm{s} < 1 < x_\textrm{l}, y_\textrm{l}$, which will be useful later.

Since the population will eventually either go extinct or reach full capacity, the generating function $\Phi_\textrm{s}(x,y)$ that corresponds to the stationary distribution $\mathbb{P}_{m,n}(t\to\infty)$ should be of the form
\begin{equation}
\Phi_\textrm{s}(x,y) = Q + (1-Q) \, \sum_{\ell=0}^{K} C_\ell \, x^\ell \, y^{K-\ell} ,
\end{equation}
where the constants $\{ C_\ell \}$ satisfy $\sum_{\ell=0}^{K} C_\ell = 1$, and $Q$ is the extinction probability, $Q = \Phi_\textrm{s}(0,0) = \mathbb{P}_{0,0}(t\to\infty)$. We have already used the normalization condition $\Phi_\textrm{s}(1,1) = 1$. Since there are not enough conserved quantities like Eq.~(\ref{eq:bc-s}), we cannot fix all constants $\{ C_\ell \}$ without actually solving the equation (\ref{eq:phi-mixed}) from initial conditions. However, an approximation of $Q$ can be obtained using Eq.~(\ref{eq:bc-s}), which implies
\begin{equation}
Q + (1-Q) \, \sum_{\ell=0}^{K} C_\ell \, x_\textrm{s}^\ell \, y_\textrm{s}^{K-\ell} = x_\textrm{s}^{m_0} \, y_\textrm{s}^{n_0} .
\end{equation}
Notice that all terms involving $\{ C_\ell \}$ are $K$-th order in $x_\textrm{s}$ and $y_\textrm{s}$, while the last term is of order $m_0 + n_0$; since $0 < x_\textrm{s}, y_\textrm{s} < 1$, for $K \gg m_0 + n_0 \geq 1$ we obtain
\begin{equation}
Q \approx x_\textrm{s}^{m_0} \, y_\textrm{s}^{n_0} .
\end{equation}

In particular, $Q \approx x_\textrm{s}$ if $(m_0, n_0) = (1, 0)$, and $y_\textrm{s}$ if $(m_0, n_0) = (0, 1)$. Therefore, $x_\textrm{s}$ is the extinction probability of a local population founded by an individual of phenotype $\textrm{A}$, and $y_\textrm{s}$ is that for phenotype $\textrm{B}$. Note that, for $q_\textrm{A} > q_\textrm{B}$, one finds $x_\textrm{s} > y_\textrm{s}$ for all $\rho$, since, intuitively, phenotype $\textrm{A}$ is more prone to demographic fluctuation that may cause extinction. For the same reason, both $x_\textrm{s}$ and $y_\textrm{s}$ increases monotonically with $\rho$, which is obvious from Eq.~(\ref{eq:xsys}). In the limit $\rho \to 1$ we have $x_\textrm{s} \to q_\textrm{A}$ as it should be, whereas $y_\textrm{s} \to q_\textrm{B}$ in the limit $\rho \to 0$.

\section{Asymptotic expansion rate $W$} \label{app:asymptotic}

Here we show how the asymptotic expansion rate $W$ is calculated. We start from the patch dynamics and derive the matrix $\boldsymbol{H}$ whose largest real eigenvalue gives $W$. We then obtain approximate expressions for $W$ in the limit of small and large dispersal rate $\mu$. We first illustrate the method by treating the simple case of a single phenotype, and then generalize to the case of a mixed strategy and of locally fluctuating environments.

\subsection{single phenotype} \label{app:asymp-single}

For a single phenotype, each local population can be labeled by its patch type $\textrm{P}_n$, where $n$ is the population size. Then the birth, death, and dispersal processes can be described by the patch dynamics
\begin{align}
\textrm{P}_n &\xrightarrow{\mathmakebox[0.3in]{\beta_n}} \textrm{P}_{n+1} , \label{eq:patch-birth1} \\
\textrm{P}_n &\xrightarrow{\mathmakebox[0.3in]{\delta_n}} \textrm{P}_{n-1} , \label{eq:patch-death1} \\
\textrm{P}_n + \textrm{P}_{n'} &\xrightarrow{\mathmakebox[0.3in]{\mu_{nn'}}} \textrm{P}_{n-1} + \textrm{P}_{n'+1} , \label{eq:patch-migration1}
\end{align}
where $n = 1, \cdots, K$; $n' = 0, \cdots, K-1$; and $n' \neq n-1$. The rate constants are $\beta_n = \beta \, n(1-n/K)$, $\delta_n = \delta \, n(1-n/K)$, and $\mu_{nn'} = \mu \, n(1-n'/K)/L$. Let $m_n$ be the number of patches $\textrm{P}_n$, and $L = \sum_{n=0}^{K} m_n$ the total number of available patches. If there is an unlimited supply of patches, $L \to \infty$, then the dispersal process (\ref{eq:patch-migration1}) reduces to
\begin{equation} \label{eq:patch-migration-unlimited1}
\textrm{P}_n \xrightarrow{\mathmakebox[0.3in]{\mu_n}} \textrm{P}_{n-1} + \textrm{P}_1 ,
\end{equation}
where $n = 2, \cdots, K$, and $\mu_n = \mu \, n$. This is because with probability $1$ a migrating individual would end up in an empty new patch. In this limit, the overall abundance of the species will asymptotically expand at a constant rate $W$, and the patch numbers will approach a steady distribution.

The deterministic dynamics of the patch numbers for processes (\ref{eq:patch-birth1}, \ref{eq:patch-death1}) and (\ref{eq:patch-migration-unlimited1}) is given by
\begin{align} \label{eq:patch-single}
\dot{m}_n &= \beta_{n-1} \, m_{n-1} +  (\delta_{n+1} + \mu_{n+1}) \, m_{n+1} \nonumber \\
&\quad - (\beta_n + \delta_n + \mu_n) \, m_n  + \delta_{n,1} \sum_{n'=1}^{K} \mu_{n'} \, m_{n'} ,
\end{align}
for $n = 1, \cdots, K$, where $m_{K+1} \equiv 0$. This can be written in a matrix form as $\dot{\boldsymbol{m}} = \boldsymbol{H} \cdot \boldsymbol{m}$, where $\boldsymbol{m} \equiv (m_1, \cdots, m_K)^T$ and
\begin{align} \label{eq:Hmat}
&\boldsymbol{H} \equiv \\
&\left( \begin{array}{ccccc}
- \beta_1 \! - \! \delta_1   & \!\!\!\! \delta_2 \! + \! 2 \, \mu_2               & \! \mu_3                  & \! \cdots         & \! \mu_K \\
\beta_1                      & \!\!\!\! - \beta_2 \! - \! \delta_2 \! - \! \mu_2  & \! \delta_3 \! + \! \mu_3 & \!                & \! \\
                             & \!\!\!\! \ddots                                    & \! \ddots                 & \! \ddots         & \! \\
                             & \!\!\!\!                                           & \! \ddots                 & \! \ddots         & \! \delta_K \! + \! \mu_K \\
                             & \!\!\!\!                                           & \!                        & \! \beta_{K-1}    & \! - \beta_K \! - \! \delta_K \! - \! \mu_K
\end{array} \right) \! . \nonumber
\end{align}
The asymptotic expansion rate, $W$, is given by the largest real eigenvalue $\lambda^{(1)}$ of the matrix $\boldsymbol{H}$, and the steady patch number distribution is given by the corresponding right eigenvector $\boldsymbol{\xi}^{(1)}$, normalized such that $\sum_{n=1}^{K} \xi^{(1)}_n = 1$.

Analytic expressions of $W$ can be obtained in the limits where the dispersal rate $\mu$ is large or small. Notice that the matrix $\boldsymbol{H}$ can be separated into two parts, one independent of $\mu$ and the other proportional to $\mu$,
\begin{align} \label{eq:decomp}
\boldsymbol{H} &= \left( \begin{array}{ccccc}
-\beta_1 \! - \! \delta_1      & \delta_2                   &           &           &                   \\
\beta_1                     & -\beta_2 \! - \! \delta_2    & \delta_3       &           &                   \\
                        & \ddots                & \ddots    & \ddots    &                   \\
                        &                       & \ddots    & \ddots    & \delta_K               \\
                        &                       &           & \beta_{K-1}   & -\beta_K \! - \! \delta_K
\end{array} \right) \nonumber \\
&\quad + \left( \begin{array}{ccccc}
0   & 2 \, \mu_2  & \mu_3       & \cdots    & \mu_K   \\
    & - \mu_2     & \mu_3       &           &       \\
    &           & \ddots    & \ddots    &       \\
    &           &           & \ddots    & \mu_K   \\
    &           &           &           & - \mu_K
\end{array} \right)
\equiv \boldsymbol{H}^{\beta\delta} + \boldsymbol{H}^{\mu} .
\end{align}
When $\mu \gg \beta, \delta$, one can treat $\boldsymbol{H}^{\mu}$ as the dominant part and $\boldsymbol{H}^{\beta\delta}$ as a perturbation. The largest real eigenvalue of $\boldsymbol{H}^{\mu}$ is $\lambda^{\mu} = 0$, and its right and left eigenvectors are $\boldsymbol{\xi}^{\mu} = (1, 0, \cdots, 0)^T$ and $\boldsymbol{\eta}^{\mu} = (1, 2, \cdots, K)^T$. The first-order perturbation of the eigenvalue is given by
\begin{equation}
\Delta\lambda^{(1)} = {\boldsymbol{\eta}^{\mu}}^T \cdot \boldsymbol{H}^{\beta\delta} \cdot \boldsymbol{\xi}^{\mu} = (\beta - \delta) \, (1 - 1/K) .
\end{equation}
Therefore, assuming $K \gg 1$, the asymptotic expansion rate is approximately given by
\begin{equation} \label{eq:Wr}
W \approx r .
\end{equation}

When $\mu \ll \beta/K, \delta/K$, one can instead treat $\boldsymbol{H}^{\beta\delta}$ as the dominant part and $\boldsymbol{H}^{\mu}$ as a perturbation. Since $\beta_K = \delta_K = 0$, the largest real eigenvalue of $\boldsymbol{H}^{\beta\delta}$ is also zero, $\lambda^{\beta\delta} = 0$, and its eigenvectors are $\boldsymbol{\xi}^{\beta\delta} = (0, \cdots, 0, 1)^T$ and $\boldsymbol{\eta}^{\beta\delta} = \{ (1-q^n)/(1-q^K), n = 1,\cdots,K \}$. Hence the first-order perturbation of the eigenvalue is
\begin{equation}
\Delta\lambda^{(1)} = {\boldsymbol{\eta}^{\beta\delta}}^T \cdot \boldsymbol{H}^{\mu} \cdot \boldsymbol{\xi}^{\beta\delta} = \mu K (1-q) \, \frac{(1-q^{K-1})}{(1-q^K)} .
\end{equation}
Therefore, assuming $K \gg 1$, one finds
\begin{equation} \label{eq:Wq}
W \approx \mu K (1-q) .
\end{equation}

\subsection{mixed population} \label{app:asymp-mixed}

We will use the same method as above to treat the case of mixed populations with two phenotypes. The patch dynamics is given in the main text by Eqs.~(\ref{eq:patch-birthA}-\ref{eq:patch-migrationB}), and the deterministic equation of the patch numbers is given by Eq.~(\ref{eq:patch-mixed}), i.e.,
\begin{align}
&\dot{m}_{n,l} = \rho \, \beta_{n-1,l-1} \, m_{n-1,l-1} + (1-\rho) \, \beta_{n-1,l} \, m_{n-1,l} \nonumber \\
&+ (\gamma_{n+1,l+1} + \mu_{l+1}) \, m_{n+1,l+1} + (\delta_{n+1,l} + \mu_{n-l+1}) \, m_{n+1,l} \nonumber \\
&- (\beta_{n,l} + \gamma_{n,l} + \delta_{n,l} + \mu_n) \, m_{n,l} \nonumber \\
&+ \delta_{n,1} \, \sum_{n'=1}^{K} \sum_{l'=0}^{n'} (\delta_{l,1} \, \mu_{l'} + \delta_{l,0} \, \mu_{n'-l'}) \, m_{n',l'} ,
\end{align}
where $n = 1, \cdots, K$ and $l = 0, \cdots, n$; the rate constants are $\beta_{nl} = \big( \beta_\textrm{A} l + \beta_\textrm{B} (n-l) \big) \, (1-n/K)$, $\gamma_{nl} = \delta_\textrm{A} \, l \, (1-n/K)$, and $\delta_{nl} = \delta_\textrm{B} \, (n-l) \, (1-n/K)$. Note that $m_{n,l}$ is set to $0$ for $n = 0, K\!+\!1$ and for $l < 0$ or $l > n$. This equation can also be cast in a matrix form for a vector $\boldsymbol{m}$ with composite indices $(nl)$, i.e., $\dot{m}_{nl} = \sum_{n'l'} H_{nl,n'l'} \, m_{n'l'}$. The elements of the matrix $\boldsymbol{H}$ are
\begin{align} \label{eq:Hnl}
&H_{nl,n'l'} = \rho \, \beta_{n',l'} \, \delta_{n',n-1} \, \delta_{l',l-1} + (1-\rho) \, \beta_{n',l'} \, \delta_{n',n-1} \, \delta_{l',l} \nonumber \\
&+ (\gamma_{n',l'} + \mu_{l'}) \delta_{n',n+1} \, \delta_{l',l+1} + (\delta_{n',l'} + \mu_{n'-l'}) \delta_{n',n+1} \, \delta_{l',l} \nonumber \\
&- (\beta_{n',l'} + \gamma_{n',l'} + \delta_{n',l'} + \mu_{n'}) \, \delta_{n',n} \, \delta_{l',l} \nonumber \\
&+ \delta_{n,1} (\delta_{l,1} \, \mu_{l'} + \delta_{l,0} \, \mu_{n'-l'}) .
\end{align}
The asymptotic expansion rate $W$ is given by the largest real eigenvalue of this matrix $\boldsymbol{H}$.

Like for Eq.~(\ref{eq:decomp}), $\boldsymbol{H}$ can be decomposed as $\boldsymbol{H} = \boldsymbol{H}^{\beta\delta} + \boldsymbol{H}^\mu$, where
\begin{align}
H^{\beta\delta}_{nl,n'l'} &= \rho \, \beta_{n',l'} \, \delta_{n',n-1} \, \delta_{l',l-1} + (1 \!-\! \rho) \beta_{n',l'} \, \delta_{n',n-1} \, \delta_{l',l} \nonumber \\
&\quad + \gamma_{n',l'} \, \delta_{n',n+1} \, \delta_{l',l+1} + \delta_{n',l'} \, \delta_{n',n+1} \, \delta_{l',l} \nonumber \\
&\quad - (\beta_{n',l'} + \gamma_{n',l'} + \delta_{n',l'}) \, \delta_{n',n} \, \delta_{l',l} , \\
H^{\mu}_{nl,n'l'} &= \mu_{l'} \, \delta_{n',n+1} \, \delta_{l',l+1} + \mu_{n'-l'} \, \delta_{n',n+1} \, \delta_{l',l} \\
&\quad - \mu_{n'} \, \delta_{n',n} \, \delta_{l',l} + \delta_{n,1} (\delta_{l,1} \, \mu_{l'} + \delta_{l,0} \, \mu_{n'-l'}) . \nonumber
\end{align}
Thus, $\boldsymbol{H}^{\beta\delta}$ is independent of $\mu$, and $\boldsymbol{H}^{\mu}$ is proportional to $\mu$. In the limit of large or small $\mu$, the largest real eigenvalue of $\boldsymbol{H}$ can be calculated as follows.

For large $\mu$, we again treat $\boldsymbol{H}^{\mu}$ as the dominant part and $\boldsymbol{H}^{\beta\delta}$ as a perturbation. The largest real eigenvalue of $\boldsymbol{H}^{\mu}$ is $0$, which turns out to be degenerate with degree 2. The right eigenvectors corresponding to this eigenvalue are $\xi^{(1)}_{nl} = \delta_{n,1} \delta_{l,1}$ and $\xi^{(2)}_{nl} = \delta_{n,1} \delta_{l,0}$. Since the leading eigenvector of $\boldsymbol{H}$ represents the steady patch number distribution, it means that the leading contribution to the steady distribution comes from patches that are occupied by just one individual, $n = 1$. This agrees with our intuition that, in the limit of frequent dispersal, individuals are scattered out over many patches.

The left eigenvectors that are orthonormal to $\boldsymbol{\xi}^{(1)}$ and $\boldsymbol{\xi}^{(2)}$ are $\eta^{(1)}_{nl} = l$ and $\eta^{(2)}_{nl} = n-l$. Therefore, the first-order perturbation of the degenerate eigenvalues are given by the $2 \times 2$ matrix
\begin{align} \label{eq:HAB}
\Delta \boldsymbol{H}^\mu &= \left( \begin{array}{c} {\boldsymbol{\eta}^{(1)}}^T \\ {\boldsymbol{\eta}^{(2)}}^T \end{array} \right)
\cdot \boldsymbol{H}^{\beta\delta} \cdot
\left( \begin{array}{cc} \boldsymbol{\xi}^{(1)} & \boldsymbol{\xi}^{(2)} \end{array} \right) \\
&= \Big( 1 - \frac{1}{K} \Big) \, \left( \begin{array}{cc}
\rho \, \beta_\textrm{A} - \delta_\textrm{A} , & \rho \, \beta_\textrm{B} \\
(1-\rho) \, \beta_\textrm{A} , & (1-\rho) \beta_\textrm{B} - \delta_\textrm{B}
\end{array} \right) . \nonumber
\end{align}
It may be recognized that, for $K \to \infty$, this is exactly the growth matrix that governs the deterministic dynamics of a single mixed population (see Eqs.~(\ref{eq:birth-mixed},\ref{eq:death-mixed})),
\begin{align}
\dot{n}_\textrm{A} &= (\rho \, \beta_\textrm{A} - \delta_\textrm{A}) \, n_\textrm{A} + \rho \, \beta_\textrm{B} \, n_\textrm{B} , \\
\dot{n}_\textrm{B} &= (1-\rho) \, \beta_\textrm{A} \, n_\textrm{A} + ((1-\rho) \beta_\textrm{B} - \delta_\textrm{B}) \, n_\textrm{B} .
\end{align}
The growth rate of this mixed population is given by the larger eigenvalue of the growth matrix,
\begin{align} \label{eq:rm}
&r_\textrm{m} (\rho) = \frac{1}{2} \bigg[ \big( \rho \, \beta_\textrm{A} - \delta_\textrm{A} + (1-\rho) \beta_\textrm{B} - \delta_\textrm{B} \big) \\
&+ \sqrt{ \big( \rho \beta_\textrm{A} \! - \! \delta_\textrm{A} \! - \! (1 \! - \! \rho) \beta_\textrm{B} \! + \! \delta_\textrm{B} \big)^2 + 4 \rho (1-\rho) \beta_\textrm{A} \beta_\textrm{B} } \, \bigg] . \nonumber
\end{align}
Therefore, similar to Eq.~(\ref{eq:Wr}), we recover the relation $W \approx r_\textrm{m}(\rho)$, as quoted in the main text.

One can easily check that $r_\textrm{m}(0) = r_\textrm{B}$ and $r_\textrm{m}(1) = r_\textrm{A}$, where $r_a \equiv \beta_a - \delta_a$ for each phenotype $a = \textrm{A}, \textrm{B}$. For $0 < \rho < 1$, the value of $r_\textrm{m}(\rho)$ lies in between $r_\textrm{B}$ and $r_\textrm{A}$. Indeed, for $r_\textrm{A} > r_\textrm{B}$, one finds that $r_\textrm{m}(\rho)$ increases monotonically with $\rho$, since
\begin{align}
r_\textrm{m}'(\rho) &= \frac{(r_\textrm{A} - r_\textrm{B}) r_\textrm{m} + (r_\textrm{m} - r_\textrm{B}) \delta_\textrm{A} + (r_\textrm{A} - r_\textrm{m}) \delta_\textrm{B}}{\sqrt{ \big( \rho \beta_\textrm{A} \! - \! \delta_\textrm{A} \! - \! (1 \! - \! \rho) \beta_\textrm{B} \! + \! \delta_\textrm{B} \big)^2 + 4 \rho (1-\rho) \beta_\textrm{A} \beta_\textrm{B} }} \nonumber \\
&> 0 .
\end{align}

For small $\mu$, $\boldsymbol{H}^{\beta\delta}$ becomes the dominant part of $\boldsymbol{H}$, and $\boldsymbol{H}^{\mu}$ the perturbation. Since $\beta_{K,l} = \delta_{K,l} = 0$ for all $l = 0, \cdots, K$, the largest real eigenvalue of $\boldsymbol{H}^{\beta\delta}$, $\lambda^{(1)} = 0$, is $(K+1)$-degree degenerate. The corresponding right eigenvectors are simply $\xi^{(s)}_{nl} = \delta_{n,K} \delta_{l,s}$, where $s = 0, \cdots, K$. Unfortunately, not all left eigenvectors can be found analytically (there is one, $\eta_{nl} \propto 1 - x_\textrm{s}^l \, y_\textrm{s}^{n-l}$, where $x_\textrm{s}$, $y_\textrm{s}$ are given in Sec.~\ref{app:ex-mixed}). Therefore, we will resort to a different method based on the following approximation.

As shown by the eigenvectors $\xi^{(s)}_{nl}$, in the limit of rare dispersal, most occupied patches are full, $n = K$. Since the timescale of birth and death processes within a patch is much shorter than the dispersal timescale, in terms of the latter, newly occupied patches either go extinct immediately or quickly grow to full capacity. Therefore, we may ignore the partially filled patches and consider only the full patches. For $K \gg 1$, the phenotype composition of a full patch is approximately given by the eigenvector of the matrix (\ref{eq:HAB}) corresponding to the eigenvalue $r_\textrm{m}$, $\boldsymbol{\xi}_\textrm{m}^{(1)} = (\xi_\textrm{A}, \xi_\textrm{B})^T$, where
\begin{equation}
\xi_\textrm{A} \equiv \frac{r_\textrm{m} - r_\textrm{B}}{r_\textrm{A} - r_\textrm{B}} , \quad \xi_\textrm{B} \equiv \frac{r_\textrm{A} - r_\textrm{m}}{r_\textrm{A} - r_\textrm{B}} .
\end{equation}

Therefore, for a given full patch, the rate of emigration of an individual with phenotype $\textrm{A}$ is $\mu K \xi_\textrm{A}$, and that of an individual with phenotype $\textrm{B}$ is $\mu K \xi_\textrm{B}$. Next we have to take into account the chance of successful colonization of a new patch by the emigrated individual. By the results from Sec.~\ref{app:ex-mixed}, the extinction probability of a local population founded by an individual of phenotype $\textrm{A}$ is $x_\textrm{s}$, and that for phenotype $\textrm{B}$ is $y_\textrm{s}$. Therefore, the rate of successful colonization is
\begin{equation}
W \approx \mu K \xi_\textrm{A} (1 - x_\textrm{s}) + \mu K \xi_\textrm{B} (1 - y_\textrm{s}) = \mu K (1 - q_\textrm{m}) ,
\end{equation}
which has the same form as Eq.~(\ref{eq:Wq}), with an average extinction probability
\begin{equation} \label{eq:qm}
q_\textrm{m}(\rho) \equiv \frac{r_\textrm{m} - r_\textrm{B}}{r_\textrm{A} - r_\textrm{B}} \, x_\textrm{s} + \frac{r_\textrm{A} - r_\textrm{m}}{r_\textrm{A} - r_\textrm{B}} \, y_\textrm{s} .
\end{equation}
One readily checks that $q_\textrm{m}(1) = q_\textrm{A}$ and $q_\textrm{m}(0) = q_\textrm{B}$. For $r_\textrm{A} > r_\textrm{B}$ and $q_\textrm{A} > q_\textrm{B}$, $q_\textrm{m}(\rho)$ monotonically increases with $\rho$, since
\begin{equation}
q_\textrm{m}'(\rho) = \frac{x_\textrm{s} - y_\textrm{s}}{r_\textrm{A} - r_\textrm{B}} \, r_\textrm{m}'(\rho) + \frac{r_\textrm{m} - r_\textrm{B}}{r_\textrm{A} - r_\textrm{B}} \, x_\textrm{s}'(\rho) + \frac{r_\textrm{A} - r_\textrm{m}}{r_\textrm{A} - r_\textrm{B}} \, y_\textrm{s}'(\rho) ,
\end{equation}
where each term is positive.

\subsection{local environmental variations} \label{app:asymp-flu}

We now consider the case where the local environment of each patch can switch between two conditions $\textrm{X}$ and $\textrm{Y}$. Denote a patch with local environment $\varepsilon = \textrm{X}$ or $\textrm{Y}$ by $\textrm{P}^\varepsilon$, and classify the patches by their local population size and composition as before, denoted by subscripts $(nl)$. Then the patch dynamics can be written as:
\begin{align}
& \textrm{P}^\varepsilon_{n,l} \xrightarrow{\mathmakebox[0.6in]{\rho \, \beta^{(\varepsilon)}_{nl}}} \textrm{P}^\varepsilon_{n+1,l+1} \; , \label{eq:env-birthA} \\
& \textrm{P}^\varepsilon_{n,l} \xrightarrow{\mathmakebox[0.6in]{(1-\rho) \beta^{(\varepsilon)}_{nl}}} \textrm{P}^\varepsilon_{n+1,l} \; , \label{eq:env-birthB} \\
& \textrm{P}^\varepsilon_{n,l} \xrightarrow{\mathmakebox[0.6in]{\gamma^{(\varepsilon)}_{nl}}} \textrm{P}^\varepsilon_{n-1,l-1} \; , \label{eq:env-deathA} \\
& \textrm{P}^\varepsilon_{n,l} \xrightarrow{\mathmakebox[0.6in]{\delta^{(\varepsilon)}_{nl}}} \textrm{P}^\varepsilon_{n-1,l} \; , \label{eq:env-deathB} \\
& \textrm{P}^\varepsilon_{n,l} \xrightarrow{\mathmakebox[0.6in]{p_{\varepsilon'} \, \mu_{l}}} \textrm{P}^\varepsilon_{n-1,l-1} + \textrm{P}^{\varepsilon'}_{1,1} \; , \label{eq:env-migrationA} \\
& \textrm{P}^\varepsilon_{n,l} \xrightarrow{\mathmakebox[0.6in]{p_{\varepsilon'} \, \mu_{n-l}}} \textrm{P}^\varepsilon_{n-1,l} + \textrm{P}^{\varepsilon'}_{1,0} \; , \label{eq:env-migrationB} \\
& \textrm{P}^\textrm{X}_{n,l} \xrightarrow{\mathmakebox[0.6in]{\alpha_\textrm{Y}}} \textrm{P}^\textrm{Y}_{n,l} \; , \\
& \textrm{P}^\textrm{Y}_{n,l} \xrightarrow{\mathmakebox[0.6in]{\alpha_\textrm{X}}} \textrm{P}^\textrm{X}_{n,l} \; ,
\end{align}
where the rate constants are defined similarly to that in the previous subsection except for the extra dependence on the environment. The last two equations describe the switching of local environment from $\textrm{Y}$ to $\textrm{X}$ and from $\textrm{X}$ to $\textrm{Y}$, with rates $\alpha_\textrm{X}$ and $\alpha_\textrm{Y}$ respectively. We have again assumed an unlimited number of available patches, $L \to \infty$. A newly colonized patch will have local environment $\textrm{X}$ with probability $p_\textrm{X} = \alpha_\textrm{X} / (\alpha_\textrm{X} + \alpha_\textrm{Y}) \equiv \epsilon$, and $\textrm{Y}$ with probability $p_\textrm{Y} = (1-\epsilon)$, which are used in Eqs.~(\ref{eq:env-migrationA},\ref{eq:env-migrationB}).

As before, the asymptotic expansion rate $W$ of a species in this patchy and fluctuating environment can be obtained from the deterministic dynamics of the patch numbers. Here we give some simplified derivations.

In the limit of a large dispersal rate $\mu$, as argued in Sec.~\ref{app:asymp-mixed}, individuals are always scattered out over many patches. Since the local environment of each patch fluctuates randomly and independently, there is approximately a fraction $\epsilon$ of the patches having environment $\textrm{X}$ and $(1-\epsilon)$ of them having $\textrm{Y}$ at any given time. Therefore, the species acts as a large population in a ``well mixed'' environment, with \emph{average} birth and death rates
\begin{align}
\bar{\beta}_a &= \epsilon \, \beta_a^{(\textrm{X})} + (1-\epsilon) \, \beta_a^{(\textrm{Y})} , \label{eq:Bave} \\
\bar{\delta}_a &= \epsilon \, \delta_a^{(\textrm{X})} + (1-\epsilon) \, \delta_a^{(\textrm{Y})} , \label{eq:Dave}
\end{align}
for each phenotype $a = \textrm{A}, \textrm{B}$. The problem then becomes equivalent to the case of a uniform environment treated in the previous subsection --- in the limit of large $\mu$, $W$ is given by the growth rate $r_\textrm{m}$ in Eq.~(\ref{eq:rm}), calculated from the average birth and death rates above.

In this limit, the optimal phenotype distribution $\rho^*$ can be obtained easily --- whichever phenotype that yields a faster growth rate, $\bar{r}_a = \bar{\beta}_a - \bar{\delta}_a$, is evolutionarily more successful; i.e., $\rho^* = 0$ if $\bar{r}_\textrm{A} < \bar{r}_\textrm{B}$, and $1$ if $\bar{r}_\textrm{A} > \bar{r}_\textrm{B}$. In the example used for Fig.~\ref{fig:phase} in the main text, we have $\bar{r}_\textrm{A} = \epsilon \, r_\textrm{A} - (1-\epsilon) s_\textrm{A}$ and $\bar{r}_\textrm{B} = r_\textrm{B}$, where $r_\textrm{A} = \beta_\textrm{A}^{(\textrm{X})} - \delta_\textrm{A}^{(\textrm{X})}$, $s_\textrm{A} = \delta_\textrm{A}^{(\textrm{Y})} - \beta_\textrm{A}^{(\textrm{Y})}$, and $r_\textrm{B} = \beta_\textrm{B}^{(\textrm{X})} - \delta_\textrm{B}^{(\textrm{X})}$. Hence, there is a critical value $\epsilon_C = \frac{r_\textrm{B} + s_\textrm{A}}{r_\textrm{A} + s_\textrm{A}}$, above which $\rho^* = 1$ and below which $\rho^* = 0$, as shown in Fig.~\ref{fig:phase}.

In the limit of small $\mu$, we use an approximation similar to the one used in Sec.~\ref{app:asymp-mixed}. Consider a special case where phenotype $\textrm{A}$ is both fast-growing and better-surviving in environment $\textrm{X}$, whereas phenotype $\textrm{B}$ is fast-growing and better-surviving in $\textrm{Y}$, which will be discussed in Sec.~\ref{app:phase}. Since most occupied patches are full in this limit, we may consider only the full patches, which can have environment $\varepsilon = \textrm{X}$ or $\textrm{Y}$, denoted by $\textrm{P}^\varepsilon$. The phenotype composition of these two types of patches are approximately given by the eigenvectors of the corresponding growth matrices; let the fraction of phenotype $\textrm{B}$ in a patch $\textrm{P}^\textrm{X}$ be denoted by $\zeta_\textrm{B}$, and the fraction of phenotype $\textrm{A}$ in a patch $\textrm{P}^\textrm{Y}$ by $\zeta_\textrm{A}$, satisfying $\zeta_a \ll 1$. We assume $K \zeta_a \gg 1$, so that both phenotypes are generally present in each full patch.

Under those conditions, the patch dynamics may be reduced to the following processes:
\begin{align}
& \textrm{P}^\textrm{X} \xrightarrow{\mathmakebox[1.2in]{\mu K (1-\zeta_\textrm{B}) (1-\tilde{q}_\textrm{A}) \epsilon}} 2 \, \textrm{P}^\textrm{X} \; , \\
& \textrm{P}^\textrm{Y} \xrightarrow{\mathmakebox[1.2in]{\mu K \zeta_\textrm{A} (1-\tilde{q}_\textrm{A}) \epsilon}} \textrm{P}^\textrm{Y} + \textrm{P}^\textrm{X} \; , \\
& \textrm{P}^\textrm{X} \xrightarrow{\mathmakebox[1.2in]{\mu K \zeta_\textrm{B} (1-\tilde{q}_\textrm{B}) (1-\epsilon)}} \textrm{P}^\textrm{X} + \textrm{P}^\textrm{Y} \; , \\
& \textrm{P}^\textrm{Y} \xrightarrow{\mathmakebox[1.2in]{\mu K (1-\zeta_\textrm{A}) (1-\tilde{q}_\textrm{B}) (1-\epsilon)}} 2 \, \textrm{P}^\textrm{Y} \; , \\
& \textrm{P}^\textrm{X} \xrightarrow{\mathmakebox[1.2in]{\alpha_\textrm{Y}}} \textrm{P}^\textrm{Y} \; , \\
& \textrm{P}^\textrm{Y} \xrightarrow{\mathmakebox[1.2in]{\alpha_\textrm{X}}} \textrm{P}^\textrm{X} \; .
\end{align}
The first two equations represent processes in which an individual of phenotype $\textrm{A}$ migrates to a patch of environment $\textrm{X}$ and successfully establishes a colony; the chance of unsuccessful colonization (local extinction) is $\tilde{q}_\textrm{A} \equiv \max \{ 1, q_\textrm{A}/\rho \}$, where $q_\textrm{A} \equiv \delta_\textrm{A}^{(\textrm{X})} / \beta_\textrm{A}^{(\textrm{X})}$. Similarly, the second two equations represent an individual of phenotype $\textrm{B}$ migrating to and successfully colonizing a patch of environment $\textrm{Y}$, where $\tilde{q}_\textrm{B} \equiv \max \{ 1, q_\textrm{B}/(1-\rho) \}$ and $q_\textrm{B} \equiv \delta_\textrm{B}^{(\textrm{Y})} / \beta_\textrm{B}^{(\textrm{Y})}$. Migration of phenotype $\textrm{A}$ to environment $Y$ and phenotype $\textrm{B}$ to environment $X$ are assumed to have negligible success rates. The last two equations represent the switching of local environments; a colony would likely survive the switch if both phenotypes are present.

For the reduced processes, the deterministic dynamics of the patch numbers $m_\textrm{X}$ and $m_\textrm{Y}$ are given by
\begin{align}
\dot{m}_\textrm{X} &= \mu K (1-\zeta_\textrm{B}) (1-\tilde{q}_\textrm{A}) \epsilon \, m_\textrm{X} + \mu K \zeta_\textrm{A} (1-\tilde{q}_\textrm{A}) \epsilon \, m_\textrm{Y} \nonumber \\
&\quad - \alpha_\textrm{Y} m_\textrm{X} + \alpha_\textrm{X} m_\textrm{Y} , \\
\dot{m}_\textrm{Y} &= \mu K \zeta_\textrm{B} (1-\tilde{q}_\textrm{B}) (1-\epsilon) \, m_\textrm{X} + \mu K (1-\zeta_\textrm{A}) (1-\tilde{q}_\textrm{B}) \nonumber \\
&\quad \times (1-\epsilon) \, m_\textrm{Y} + \alpha_\textrm{Y} m_\textrm{X} - \alpha_\textrm{X} m_\textrm{Y} .
\end{align}
This can be described by a growth matrix
\begin{align}
\boldsymbol{H} &= \left( \begin{array}{cc} - \alpha_\textrm{Y} , & \alpha_\textrm{X} \\ \alpha_\textrm{Y} , & - \alpha_\textrm{X} \end{array} \right) \\
&+ \mu K \left( \begin{array}{cc}
(1-\zeta_\textrm{B}) (1-\tilde{q}_\textrm{A}) \epsilon , & \zeta_\textrm{A} (1-\tilde{q}_\textrm{A}) \epsilon \\
\zeta_\textrm{B} (1-\tilde{q}_\textrm{B}) (1-\epsilon) , & (1-\zeta_\textrm{A}) (1-\tilde{q}_\textrm{B}) (1-\epsilon)
\end{array} \right) . \nonumber
\end{align}
For small $\mu$, the leading eigenvalue which gives the asymptotic expansion rate $W$ can be calculated by treating the second term as a perturbation. If we further ignore small factors involving $\zeta_\textrm{A}$ and $\zeta_\textrm{B}$, then $W$ is approximately given by
\begin{equation} \label{eq:Wmu0ap}
W \approx \mu K \bigg[ \, \frac{\epsilon^2}{\rho} \, R(\rho - q_\textrm{A}) + \frac{(1-\epsilon)^2}{(1-\rho)} \, R(1 - \rho - q_\textrm{B}) \bigg] ,
\end{equation}
where $R(x)$ is the ramp function, $R(x) = \max\{0, x\}$. Finally, the optimal phenotype distribution $\rho^*$ can be obtained by maximizing $W$ with respect to $\rho$.

For simplicity, consider the symmetric case where $q_\textrm{A} = q_\textrm{B} \equiv q$. For $q > \frac{1}{3}$, the maximum of $W$ is reached at $\rho^* = 0$ for $\epsilon < \frac{1}{2}$, and $\rho^* = 1$ for $\epsilon > \frac{1}{2}$; hence there is no mixed strategy that is favorable. However, for $q < \frac{1}{3}$,
\begin{equation} \label{eq:rho*ap_mu0}
\rho^* = \left\{ \begin{array}{ll}
0 , & 0 \leq \epsilon < \frac{2q}{1+q} ; \\[4pt]
\epsilon , & \frac{2q}{1+q} < \epsilon < \frac{1-q}{1+q} ; \\[4pt]
1 , & \frac{1-q}{1+q} < \epsilon \leq 1 .
\end{array} \right.
\end{equation}
In this case, there is a range of intermediate $\epsilon$ values for which a mixed strategy is optimal, as shown in Fig.~\subref*{fig:strong_mu0}. Interestingly, the optimal phenotype distribution, $\rho^* = \epsilon$, being proportional to the environment distribution, is reminiscent of the ``proportional betting'' solution found for a single population in a uniformly fluctuating environment (with a slightly different setting, see \cite{Rivoire2011}).

\section{Optimal phenotype distribution $\rho^*$} \label{app:optimal}

Here we describe how the optimal phenotype distribution $\rho^*$ is calculated numerically, as for Figs.~\ref{fig:rho_opt_patchy} and \ref{fig:phase} in the main text. We also explore the range of parameters for which a bet-hedging strategy arises.

\subsection{constant environment} \label{app:opt-const}

Consider first the case where the environment is constant and the same for all patches, as described in Sec.~\ref{app:asymp-mixed}. For a given phenotype distribution $\rho$, the asymptotic expansion rate $W$ is given by the largest real eigenvalue of the matrix $\boldsymbol{H}$ in Eq.~(\ref{eq:Hnl}). $\boldsymbol{H}$ is a sparse $D \times D$ matrix, where $D$ is the number of patch types $\{ P_{nl} \}$; since $1 \leq n \leq K$ and $0 \leq l \leq n$, one finds $D = K(K+3)/2$. The eigenvalue of largest real part, $W$, and its left and right eigenvectors, $\boldsymbol{\eta}$ and $\boldsymbol{\xi}$, can be found by efficient numerical routines (e.g., the shift-invert method in ARPACK). It is checked that $W$, $\boldsymbol{\eta}$, and $\boldsymbol{\xi}$ are all real. Moreover, we may calculate the derivative of $W$ with respect to $\rho$, given by
\begin{equation}
W'(\rho) = {\boldsymbol{\eta}}^\top \cdot \Big( \frac{d\boldsymbol{H}}{d\rho} \Big) \cdot \boldsymbol{\xi} ,
\end{equation}
where $(\frac{d\boldsymbol{H}}{d\rho})$ is the constant matrix given by
\begin{equation}
\Big( \frac{d\boldsymbol{H}}{d\rho} \Big)_{nl,n'l'} = \beta_{n',l'} \, \delta_{n',n-1} \, \big( \delta_{l',l-1} - \delta_{l',l} \big) .
\end{equation}

\begin{figure*}
\centering
\includegraphics[width=\textwidth]{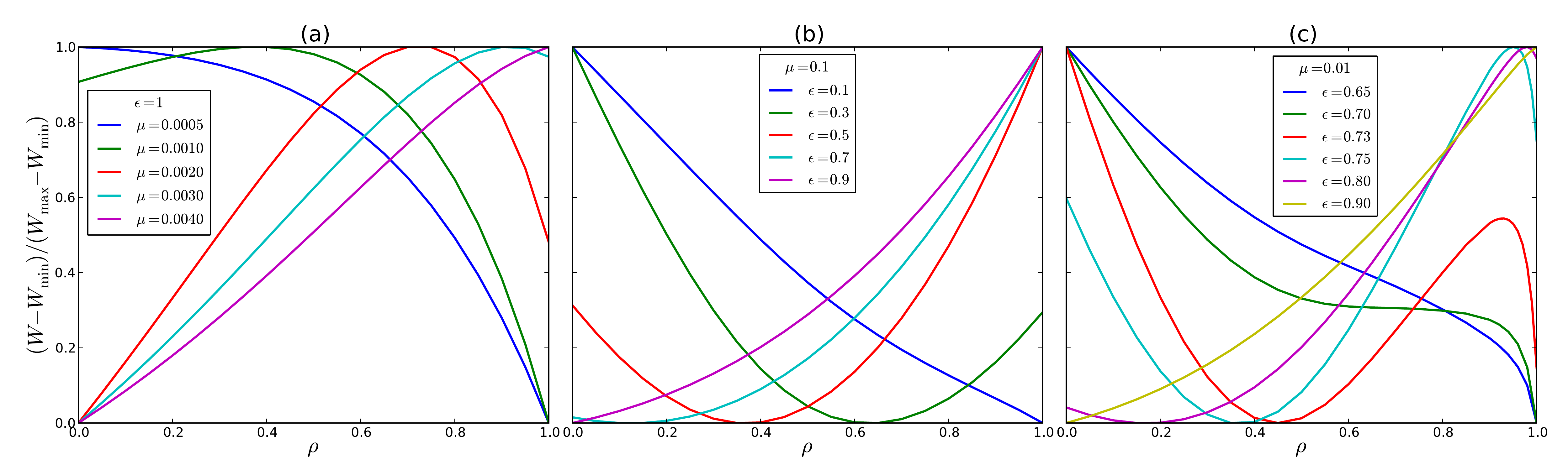}
\caption{Asymptotic expansion rate $W$ as a function of phenotype distribution $\rho$. The birth and death rates of the fast-growing phenotype $\textrm{A}$ and the better-surviving phenotype $\textrm{B}$ are the same as for Figs.~\ref{fig:rho_opt_patchy} and \ref{fig:phase} in the main text, i.e., $\beta_\textrm{A}^{(\textrm{X})} = 2$, $\delta_\textrm{A}^{(\textrm{X})} = 1$, $\beta_\textrm{A}^{(\textrm{Y})} = 0$, $\delta_\textrm{A}^{(\textrm{Y})} = 10$, $\beta_\textrm{B}^{(\textrm{X})} = \beta_\textrm{B}^{(\textrm{Y})} = 0.5$, $\delta_\textrm{B}^{(\textrm{X})} = \delta_\textrm{B}^{(\textrm{Y})} = 0.1$. The carrying capacity of each local patch is $K = 100$; the sum of the environmental switching rates is $\alpha_\textrm{X} + \alpha_\textrm{Y} = 0.1$. (\textbf{a}) the dispersal rate $\mu$ varies, while the environment is $\textrm{X}$ at all times, $\epsilon = 1$, as in Fig.~\ref{fig:rho_opt_patchy}. (\textbf{b},\textbf{c}) the environment distribution $\epsilon$ varies, while $\mu$ is fixed at $\mu = 0.1$ and $0.01$ respectively.}
\label{fig:W(rho)}
\end{figure*}

Using these methods for calculating $W$ and its derivative $W'(\rho)$, it is straight-forward to numerically maximize $W$ to find the value of $\rho^*$. Fig.~\ref{fig:W(rho)}(a) shows $W(\rho)$ for different values of $\mu$, using the same parameters ($\beta_a, \delta_a$) as for the example shown in Fig.~\ref{fig:rho_opt_patchy} of the main text. It can be seen that, for a small $\mu$, the derivative $W'(\rho)$ is negative throughout $\rho = 0$ to $1$, hence we have $\rho^* = 0$; for a large $\mu$, $W'(\rho)$ is positive at all $\rho$, hence $\rho^* = 1$, consistent with the perturbative analysis in Sec.~\ref{app:asymp-mixed}. We may define a threshold value $\mu_L$ by the condition $W'(0) = 0$, and another threshold $\mu_R$ by $W'(1) = 0$. In this example, we find $\mu_L < \mu_R$. For an intermediate value of $\mu$ between $\mu_L$ and $\mu_R$, we have $W'(0) > 0$ and $W'(1) < 0$. Therefore, $W(\rho)$ has a local maximum at $0 < \rho^* < 1$, which can be obtained by numerically finding the root of $W'(\rho) = 0$. This is the procedure used to produce Fig.~\ref{fig:rho_opt_patchy} in the main text.

\begin{figure*}
\centering
\includegraphics[width=\textwidth]{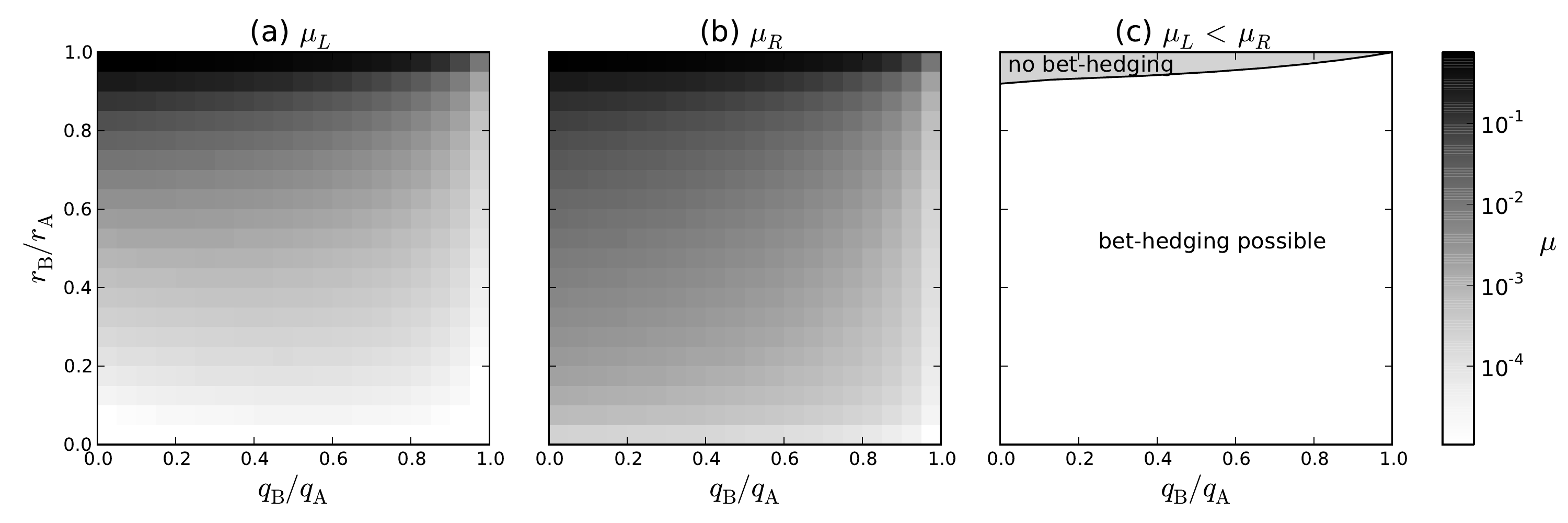}
\caption{(\textbf{a},\textbf{b}) Thresholds $\mu_L$ and $\mu_R$ defined by $W'(0) = 0$ and $W'(1) = 0$ respectively. The growth rate and the extinction risk of each phenotype $\textrm{A}, \textrm{B}$ satisfy $r_\textrm{A} > r_\textrm{B}$ and $q_\textrm{A} > q_\textrm{B}$; we set $r_\textrm{A} = 1$ by rescaling time and fix the value $q_\textrm{A} = 0.5$, then vary the values of $r_\textrm{B}$ and $q_\textrm{B}$. (\textbf{c}) Range of parameters for which $\mu_L < \mu_R$, indicating the existence of a bet-hedging regime.} \label{fig:params}
\end{figure*}

The existence of the $0 < \rho^* < 1$ phase depends on the values of the thresholds $\mu_L$ and $\mu_R$. Figs.~\ref{fig:params}(a,b) show how $\mu_L$ and $\mu_R$ vary with the parameters of the model, i.e., the growth rate $r_a$ and the extinction risk $q_a$ for each phenotype $a = \textrm{A}, \textrm{B}$. For a small range of parameters, shown in Fig.~\ref{fig:params}(c), it happens that $\mu_L > \mu_R$. This means that, for values of $\mu$ between $\mu_L$ and $\mu_R$, $W'(0) < 0$ and $W'(1) > 0$; accordingly, $W(\rho)$ develops a local minimum instead of a maximum (similar to Fig.~\ref{fig:W(rho)}(b)). In such cases, $\rho^*$ jumps discontinuously from $0$ to $1$ at a particular value of $\mu$ where $W(0) = W(1)$, missing the $0 < \rho^* < 1$ phase.

\begin{figure*}
\subfloat[\label{fig:rho_eps}]{%
  \includegraphics[width=0.5\linewidth]{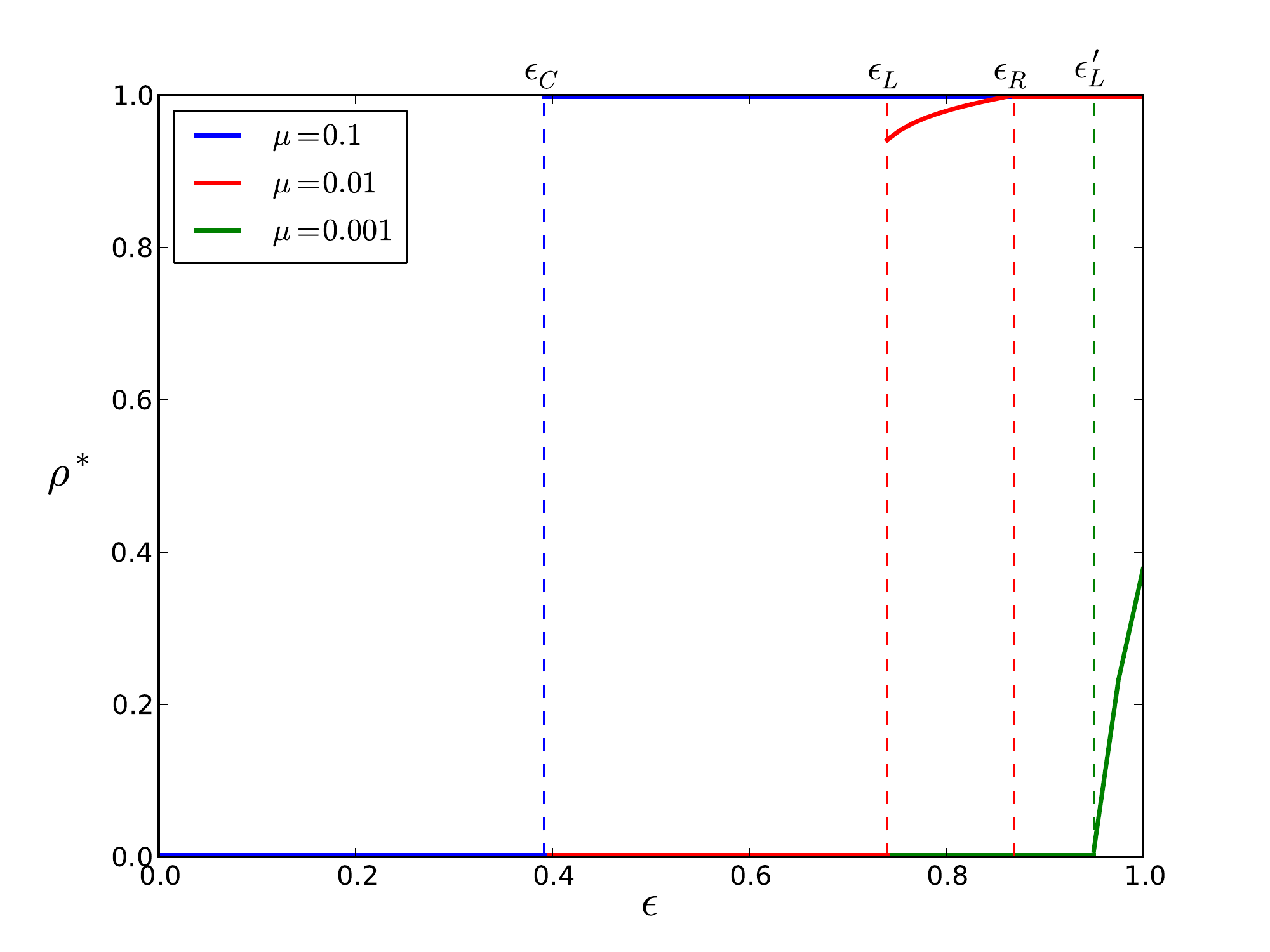}%
}\hfill
\subfloat[\label{fig:strong_mu0}]{%
  \includegraphics[width=0.49\linewidth]{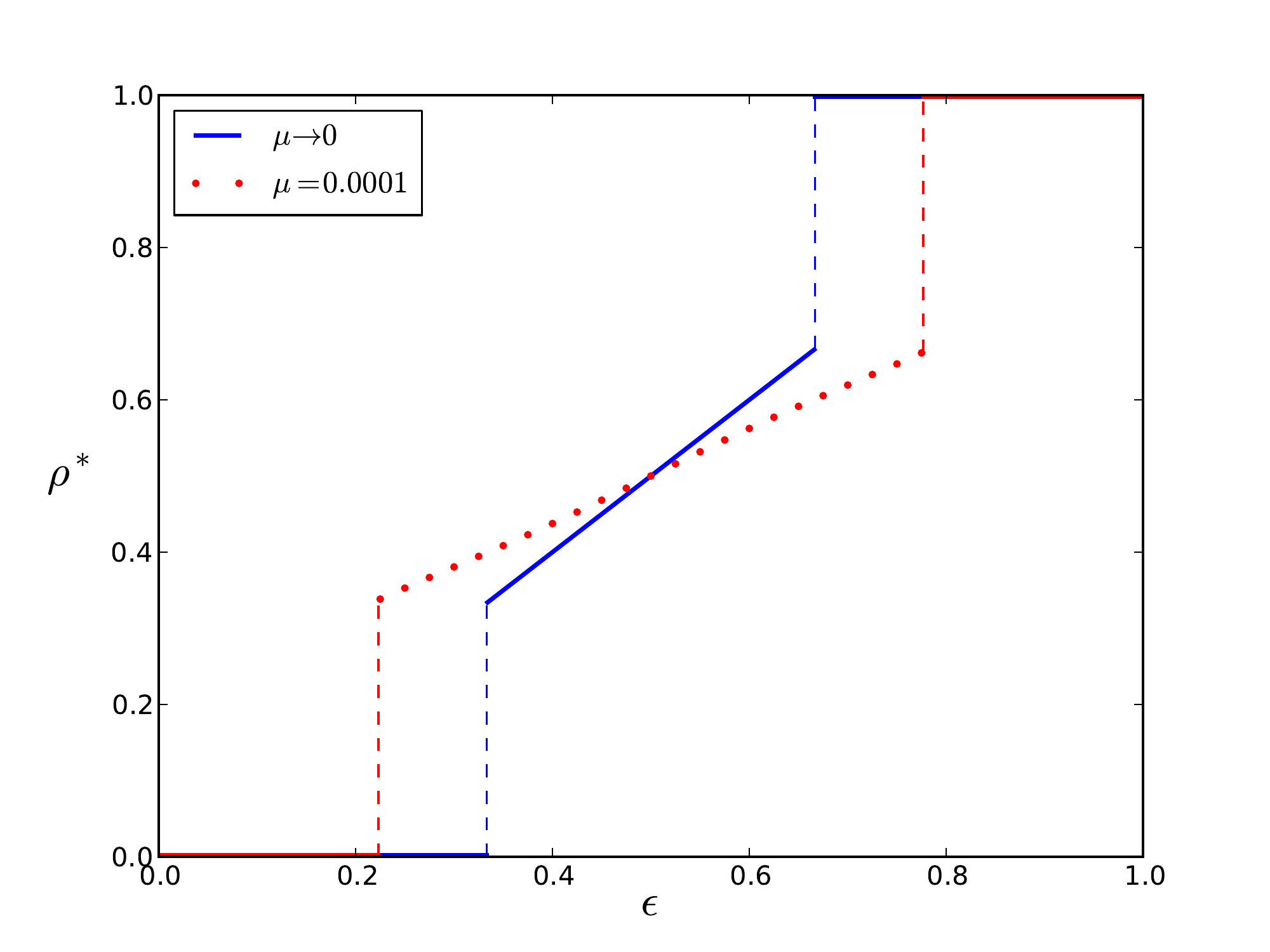}%
}
\caption{\small Optimal phenotype distribution $\rho^*$ as a function of the environment distribution $\epsilon$ for given values of the dispersal rate $\mu$. (\textbf{a}) The birth and death rates of each phenotype, the carrying capacity, and the sum of environmental switching rates are the same as for Fig.~\ref{fig:W(rho)}; the curves correspond to vertical cross-sections of the phase diagram in Fig.~\ref{fig:phase} of the main text. (\textbf{b}) The parameters are the same as for Fig.~S4(b); the continuous line is the approximate result according to Eq.~(\ref{eq:rho*ap_mu0}), and the dotted line is the exact result obtained by numerical calculations.} \label{fig:rho_opt}
\end{figure*}

\subsection{locally varying environment} \label{app:opt-local}

Generalizing to the case of locally varying environments, $W$ and $W'(\rho)$ can be numerically calculated the same way as above, except that the number of patch types, i.e., the dimension $D$ of the matrix $\boldsymbol{H}$, is doubled to account for the two types of local environments. In this case, the optimal value $\rho^*$ depends on both the dispersal rate $\mu$ and the environment distribution $\epsilon$.

Fig.~\ref{fig:W(rho)}(b) shows how $W(\rho)$ changes with $\epsilon$ for a given $\mu$. In particular, the parameters are the same as for Fig.~\ref{fig:phase} in the main text, and $\mu$ is fixed to a large value, $\mu > \mu_T$. It can be seen that $W(\rho)$ is monotonic decreasing ($W'(\rho) < 0$) for a small $\epsilon$, and becomes monotonic increasing ($W'(\rho) > 0$) for a large $\epsilon$; for some intermediate values of $\epsilon$, $W(\rho)$ has a local minimum ($W'(0) < 0$ and $W'(1) > 0$). In any case, the maximum of $W$ is reached at either $\rho^* = 0$ or $1$. Therefore, it suffices to compare the values of $W(0)$ and $W(1)$ to determine $\rho^*$. This procedure is used to produce part of Fig.~\ref{fig:phase} for $\mu > \mu_T$, where $\mu_T$ is determined below. Note that $\rho^*$ transitions from $0$ to $1$ discontinuously at a threshold value $\epsilon_C$, as shown in Fig.~\subref*{fig:rho_eps}, which is consistent with the perturbative analysis described in Sec.~\ref{app:asymp-flu} for a large $\mu$.

A more intricate situation is shown in Fig.~\ref{fig:W(rho)}(c), where $W(\rho)$ can have multiple local maxima, one at an end of the unit interval and one in the middle. This happens for certain $\mu$ values between $\mu_L$ and $\mu_T$ in Fig.~\ref{fig:phase} --- as $\epsilon$ increases from $0$ to $1$, a local maximum emerges at an intermediate value of $\rho$, even though $W'(0)$ and $W'(1)$ have the same sign. In such situations, the numerical search for $\rho^*$ is carried out as follows. In the example shown in Fig.~\ref{fig:W(rho)}(c), when both $W'(0) < 0$ and $W'(1) < 0$, the value of $W(0)$ is noted. Then one randomly searches for a value $\rho_0$ where $W'(\rho_0) > 0$; if such a $\rho_0$ is found, then one locates the local maximum $\rho^*_1$ within the range $\rho_0 < \rho^*_1 < 1$ by finding the root of $W'(\rho) = 0$. Finally, a comparison of $W(0)$ and $W(\rho^*_1)$ picks out the larger one and hence the true $\rho^*$. Such a procedure is used to produce part of Fig.~\ref{fig:phase} for $\mu_L < \mu < \mu_T$. Note that, in this example, $\rho^*$ transitions discontinuously from $0$ to a positive value $\rho^*_L$ at a certain value $\epsilon_L$, given by $W(0) = W(\rho^*_1)$; yet, as $\epsilon$ further increases, $\rho^*$ increases continuously and reaches $1$ at another value $\epsilon_R$, as shown in Fig.~\subref*{fig:rho_eps}. In particular, the point $(\mu_T, \epsilon_T)$ in Fig.~\ref{fig:phase} is determined by $\rho^*_L = 1$, i.e., when $W(0) = W(1)$ and $W'(1) = 0$. This completes the construction of Fig.~\ref{fig:phase} in the main text.

\section{Phase diagrams} \label{app:phase}

Here we consider the optimal bet-hedging strategy for a species living in a patchy and locally fluctuating environment, and give a few more distinctive examples of the phase diagram.

Assume again two possible phenotypes, $\textrm{A}$ and $\textrm{B}$, and two potential environmental conditions, $\textrm{X}$ and $\textrm{Y}$. The birth and death rates of each phenotype in a given environment are denoted by $\beta_a^{(\varepsilon)}$ and $\delta_a^{(\varepsilon)}$, where $a = \textrm{A}, \textrm{B}$, and $\varepsilon = \textrm{X}, \textrm{Y}$. The dispersal rate is $\mu$ regardless of the phenotype and the environment. The environment switches from $\textrm{Y}$ to $\textrm{X}$ and from $\textrm{X}$ to $\textrm{Y}$ with rates $\alpha_\textrm{X}$ and $\alpha_\textrm{Y}$ respectively; thus the stationary distribution of the environment is $p_\textrm{X} = \epsilon = \alpha_\textrm{X} / (\alpha_\textrm{X} + \alpha_\textrm{Y})$ and $p_\textrm{Y} = 1-\epsilon$. We assume that a newborn individual, regardless of its parent's phenotype, has probability $\rho$ of having phenotype $\textrm{A}$, and $(1-\rho)$ of having phenotype $\textrm{B}$.

\subsection{driven by demographic fluctuation} \label{app:phase-dem}

As discussed in the main text, the topology of the phase diagram for the optimal phenotype distribution $\rho^*$ can be largely determined by interpolating the behavior of $\rho^*$ between the limits $\epsilon \to 0$ and $1$, i.e., the optimal phenotype distribution in each environment alone. For a fast-growing phenotype $\textrm{A}$ and a better-surviving phenotype $\textrm{B}$ (i.e., $r_\textrm{A} > r_\textrm{B}$ and $q_\textrm{A} > q_\textrm{B}$), there is a range of dispersal rates $\mu$ between two thresholds $\mu_L$ and $\mu_R$ for which a bet-hedging strategy is optimal. As shown in the previous section, the values of $\mu_L$ and $\mu_R$ tend to vary smoothly with the parameters $r_a$ and $q_a$. Therefore, we expect the phase diagram to also transform smoothly as the parameters change.

\begin{figure*}
\subfloat[\label{fig:phase2}]{%
  \includegraphics[width=0.49\linewidth]{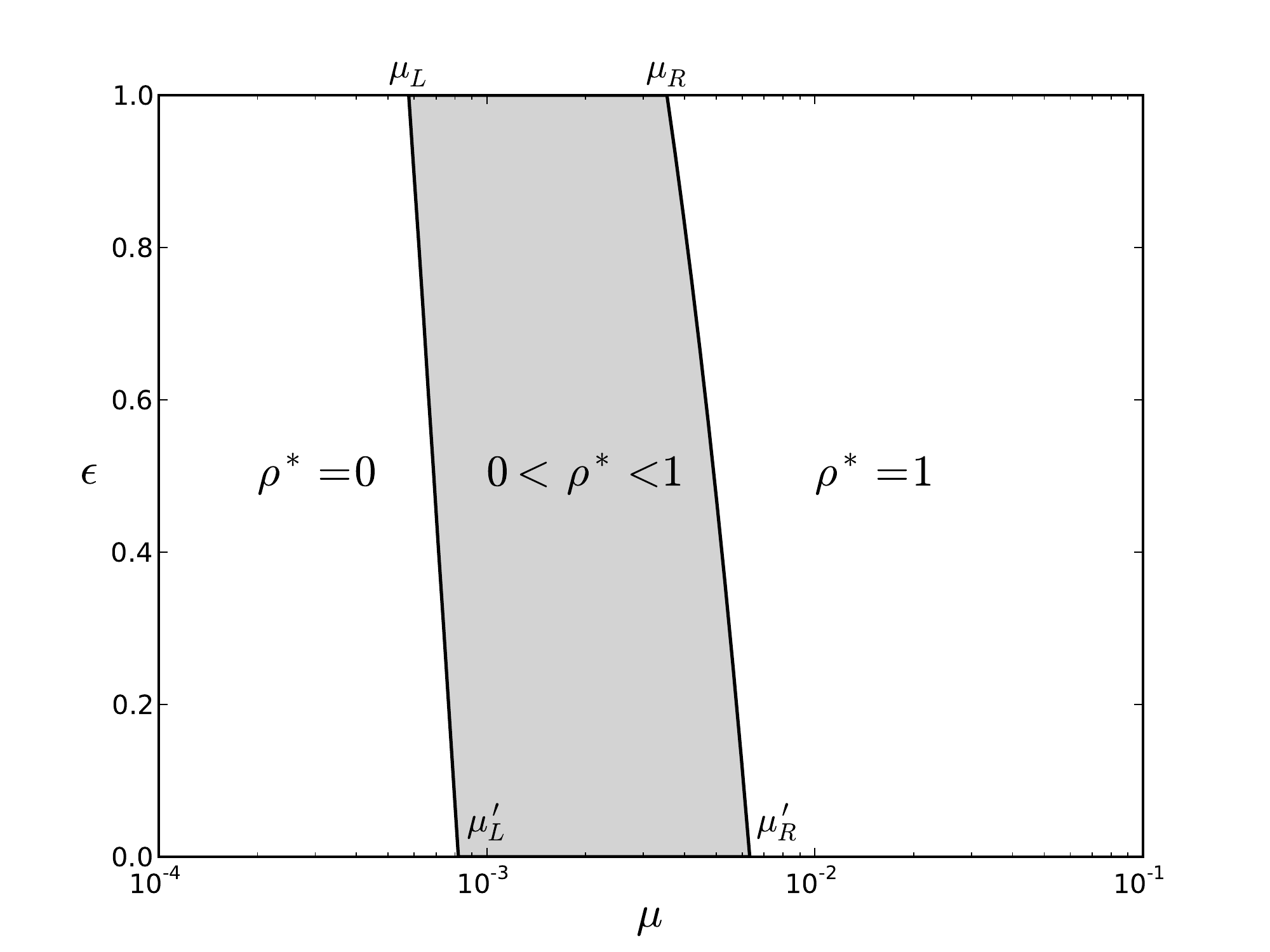}%
}\hfill
\subfloat[\label{fig:phase3}]{%
  \includegraphics[width=0.49\linewidth]{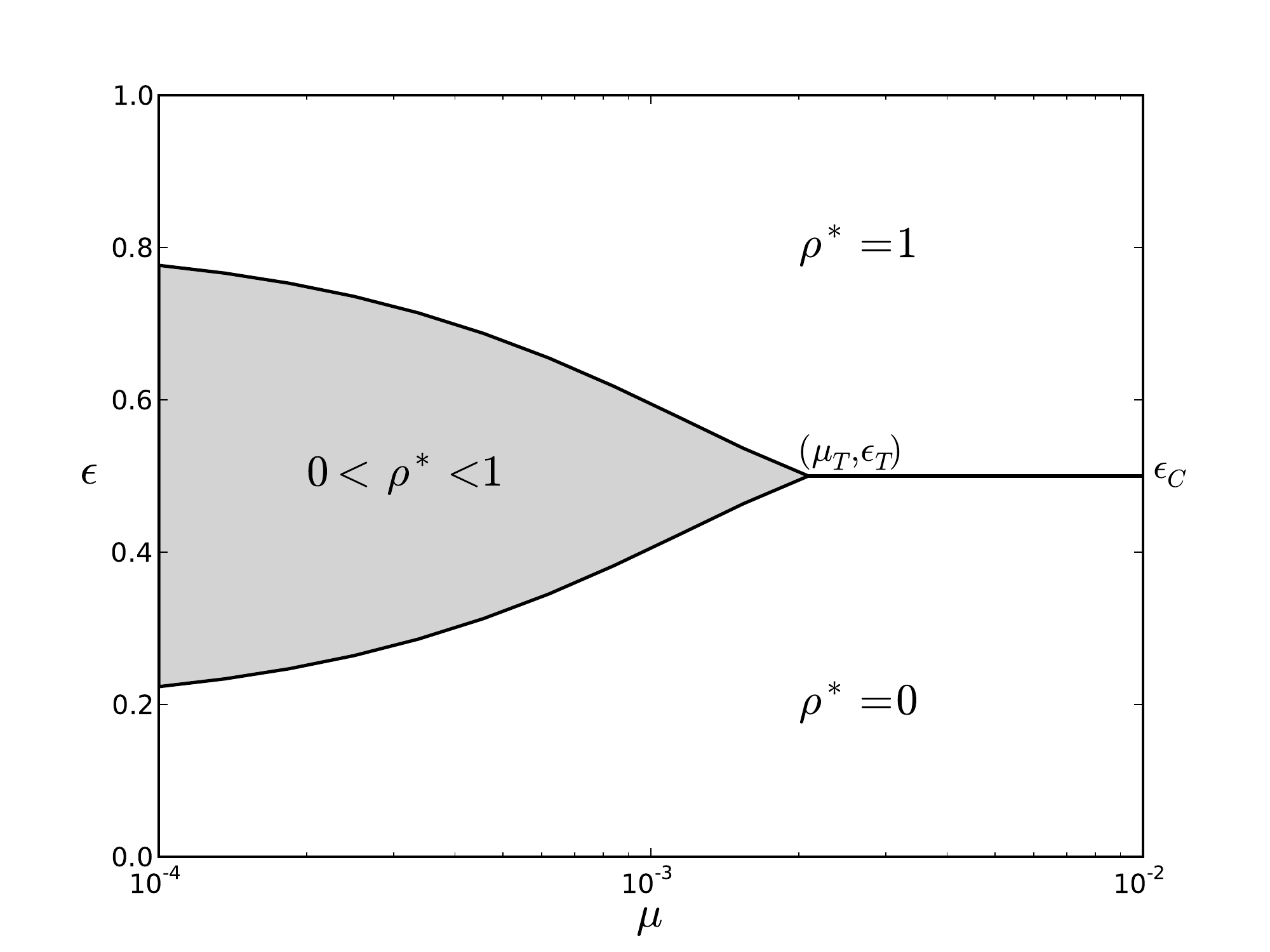}%
}
\caption{Phase diagram of the optimal phenotype distribution $\rho^*$ with respect to the dispersal rate $\mu$ and the environment distribution $p$. Shaded region marks a $0 < \rho^* < 1$ phase in which a bet-hedging strategy offers the maximum asymptotic expansion rate for the species. The carrying capacity of each local patch is $K = 100$; the sum of the environmental switching rates is $\alpha_\textrm{X} + \alpha_\textrm{Y} = 0.1$. (\textbf{a}) phenotype $\textrm{A}$ is fast-growing in both environments $\textrm{X}$ and $\textrm{Y}$, and phenotype $\textrm{B}$ is better-surviving in both $\textrm{X}$ and $\textrm{Y}$; the birth and death rates are $\beta_\textrm{A}^{(\textrm{X})} = 2$, $\delta_\textrm{A}^{(\textrm{X})} = 1$, $\beta_\textrm{A}^{(\textrm{Y})} = 3$, $\delta_\textrm{A}^{(\textrm{Y})} = 1.8$, $\beta_\textrm{B}^{(\textrm{X})} = \beta_\textrm{B}^{(\textrm{Y})} = 0.5$, $\delta_\textrm{B}^{(\textrm{X})} = \delta_\textrm{B}^{(\textrm{Y})} = 0.1$. (\textbf{b}) phenotype $\textrm{A}$ offers both a faster growth rate and a lower extinction risk in the environment $\textrm{X}$, and so does phenotype $\textrm{B}$ in the environment $\textrm{Y}$; the birth and death rates are $\beta_\textrm{A}^{(\textrm{X})} = \beta_\textrm{B}^{(\textrm{Y})} = 5$, $\delta_\textrm{A}^{(\textrm{X})} = \delta_\textrm{B}^{(\textrm{Y})} = 1$, $\beta_\textrm{A}^{(\textrm{Y})} = \beta_\textrm{B}^{(\textrm{X})} = 0$, $\delta_\textrm{A}^{(\textrm{Y})} = \delta_\textrm{B}^{(\textrm{X})} = 50$.}
\label{fig:phases}
\end{figure*}

As an example, let us see how the phase diagram in Fig.~\ref{fig:phase} of the main text can be transformed. Consider the case where the fast-growing phenotype $\textrm{A}$ in environment $\textrm{X}$ remains fast-growing in $\textrm{Y}$, and the better-surviving phenotype $\textrm{B}$ remains better-surviving. In this case, for each environment $\textrm{X}$ or $\textrm{Y}$ ($\epsilon \to 1$ or $0$), $\rho^*$ as a function of $\mu$ looks similar to Fig.~\ref{fig:rho_opt_patchy} in the main text. By interpolation, the phase diagram of $\rho^*(\mu,\epsilon)$ looks typically like in Fig.~\subref*{fig:phase2}. There is a $0 < \rho^* < 1$ phase that forms a vertical band across the top ($\epsilon = 1$) and bottom ($\epsilon = 0$) edges of the phase diagram. Hence, for any environment distribution $\epsilon$, there is a range of $\mu$ values that allow an optimal bet-hedging strategy. Such optimal bet-hedging strategies arise mainly as a result of demographic fluctuations.

\subsection{driven by environmental variation} \label{app:phase-env}

Now, consider an opposite situation where the two phenotypes are each suited for a different environment. In particular, each phenotype, $\textrm{A}$ or $\textrm{B}$, grows normally in one environment, $\textrm{X}$ or $\textrm{Y}$ respectively, yet dies quickly in the other environment. Thus, phenotype $\textrm{A}$ is both fast-growing and better-surviving in environment $\textrm{X}$, whereas phenotype $\textrm{B}$ is fast-growing and better-surviving in $\textrm{Y}$. For example, the birth and death rates are such that $\delta_\textrm{A}^{(\textrm{Y})} \gg \beta_\textrm{A}^{(\textrm{X})} > \delta_\textrm{A}^{(\textrm{X})} > \beta_\textrm{A}^{(\textrm{Y})} = 0$, and similarly $\delta_\textrm{B}^{(\textrm{X})} \gg \beta_\textrm{B}^{(\textrm{Y})} > \delta_\textrm{B}^{(\textrm{Y})} > \beta_\textrm{B}^{(\textrm{X})} = 0$. We will denote $r_\textrm{A} \equiv \beta_\textrm{A}^{(\textrm{X})} - \delta_\textrm{A}^{(\textrm{X})}$, $r_\textrm{B} \equiv \beta_\textrm{B}^{(\textrm{Y})} - \delta_\textrm{B}^{(\textrm{Y})}$; $q_\textrm{A} \equiv \delta_\textrm{A}^{(\textrm{X})} / \beta_\textrm{A}^{(\textrm{X})}$, $q_\textrm{B} \equiv \delta_\textrm{B}^{(\textrm{Y})} / \beta_\textrm{B}^{(\textrm{Y})}$; and $s_\textrm{A} \equiv \delta_\textrm{A}^{(\textrm{Y})} - \beta_\textrm{A}^{(\textrm{Y})}$, $s_\textrm{B} \equiv \delta_\textrm{B}^{(\textrm{X})} - \beta_\textrm{B}^{(\textrm{X})}$.

The phase diagram in this case can be determined again by examining the limits $\epsilon = 0, 1$ and $\mu \to 0, \infty$. First, when the environment is $\textrm{X}$ or $\textrm{Y}$ at all times, only individuals with the matching phenotype survive; hence we have $\rho^* = 0$ for $\epsilon = 0$, and $\rho^* = 1$ for $\epsilon = 1$, valid for all $\mu$. Then, in the limit $\mu \to \infty$ (see Sec.~\ref{app:asymp-flu}), the favorable phenotype is the one with a larger average growth rate among
\begin{align}
\bar{r}_\textrm{A} &= \epsilon \, r_\textrm{A} - (1-\epsilon) s_\textrm{A} , \\
\bar{r}_\textrm{B} &= - \epsilon \, s_\textrm{B} + (1-\epsilon) r_\textrm{B} .
\end{align}
Therefore, $\rho^* = 1$ for $\epsilon > \epsilon_C$, and $\rho^* = 0$ for $\epsilon < \epsilon_C$, where $\epsilon_C = \frac{r_\textrm{B} + s_\textrm{A}}{r_\textrm{A} + r_\textrm{B} + s_\textrm{A}+ s_\textrm{B}}$. Finally, in the limit $\mu \to 0$, the asymptotic behavior of $\rho^*$ can be found using an approximation described in Sec.~\ref{app:asymp-flu} --- when $q_\textrm{A}$ and $q_\textrm{B}$ are small, there is a range of intermediate $\epsilon$ values for which a bet-hedging strategy with $0 < \rho^* < 1$ is optimal. This means that, in the phase diagram of $\rho^*(\mu, \epsilon)$, the far left region is horizontally divided into three phases, $\rho^* = 0$, $0 < \rho^* < 1$, and $\rho^* = 1$, as shown in Fig.~\subref*{fig:phase3}. The phase boundaries represent discontinuous transitions in $\rho^*$ (see Fig.~\subref*{fig:strong_mu0}). As $\mu$ increases, the two boundaries extend smoothly to the right, and eventually join at a point $(\mu_T, \epsilon_T)$. Beyond this point, there is only one boundary separating the $\rho^* = 0$ and $\rho^* = 1$ phases, which becomes asymptotically horizontal as $\mu \to \infty$.

This phase diagram has a different character as compared to Fig.~\subref*{fig:phase2} (or Fig.~\ref{fig:phase} in the main text). There is no phase transition on the edges of $\epsilon = 0$ or $1$, since, for a given environment, one phenotype is advantageous in terms of both a faster growth rate and a lower extinction risk. The existence of an optimal bet-hedging strategy (the $0 < \rho^* < 1$ phase) is mainly due to environmental variations and the fact that different phenotypes are favorable in different environments, just like for a single population in a uniformly fluctuating environment. Note, however, that in the latter case a favorable bet-hedging strategy always exists for all $\epsilon$, which is not true for the patchy environment.

\section{Parent-dependent phenotypic switching} \label{app:parent}

Here we consider a generalization of our model presented in the main text, where an individual's probability of expressing a particular phenotype may depend on the phenotype of its parent. Let $a$ be the phenotype of the parent, and $\pi_{ba}$ be the probability that the offspring has phenotype $b$. For two phenotypes $\textrm{A}$ and $\textrm{B}$, we may parametrize $\pi_{ba}$ by the phenotypic switching probabilities, $\pi_\textrm{BA} \equiv \sigma_\textrm{A}$ and $\pi_\textrm{AB} \equiv \sigma_\textrm{B}$, hence $\pi_\textrm{AA} = 1 - \sigma_\textrm{A}$ and $\pi_\textrm{BB} = 1 - \sigma_\textrm{B}$, where $0 \leq \sigma_a \leq 1$.

\begin{figure}
\subfloat[\label{fig:sigma_opt}]{%
  \includegraphics[width=0.5\textwidth]{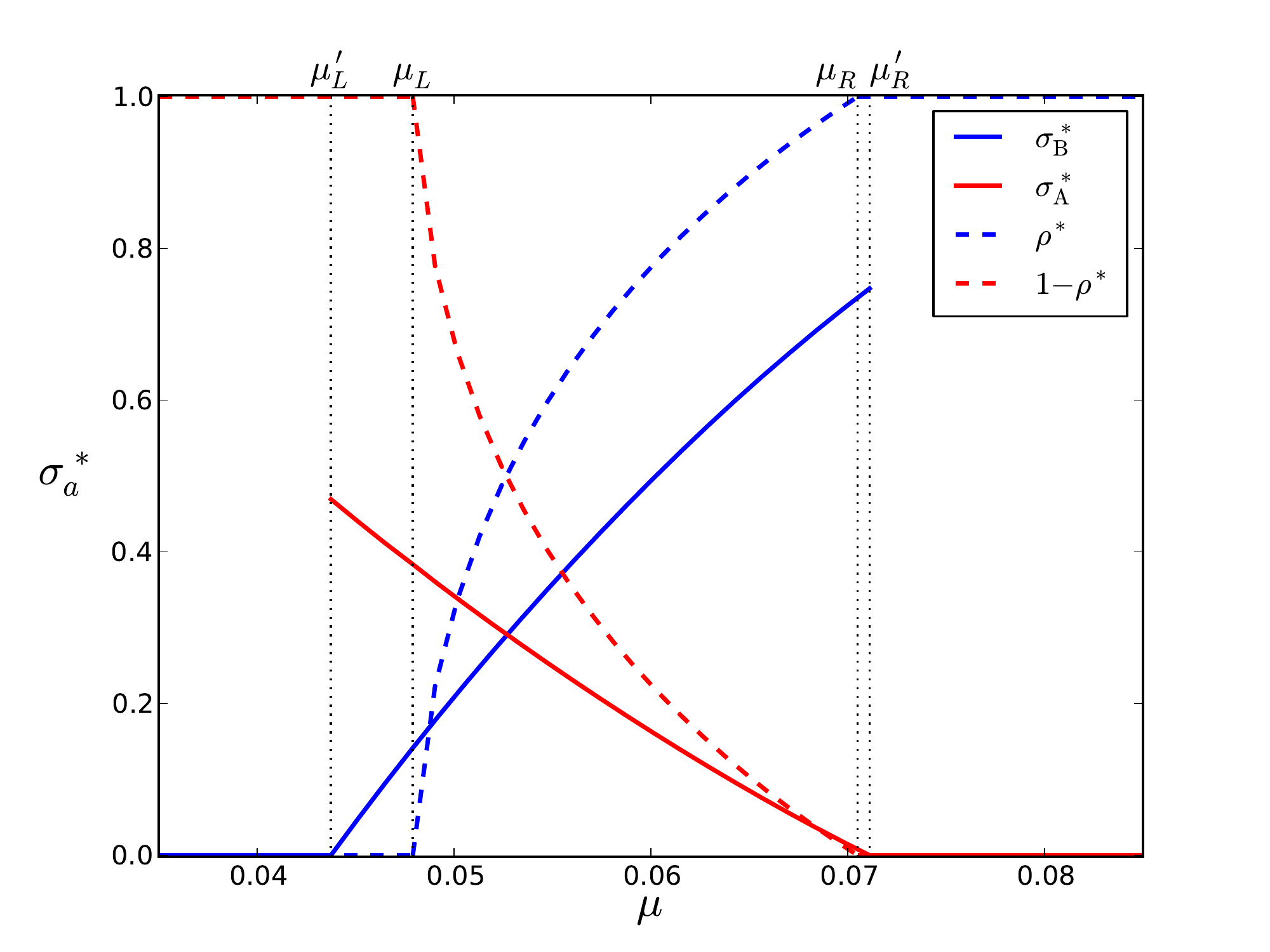}%
}\\
\subfloat[\label{fig:W_opt}]{%
  \includegraphics[width=0.5\textwidth]{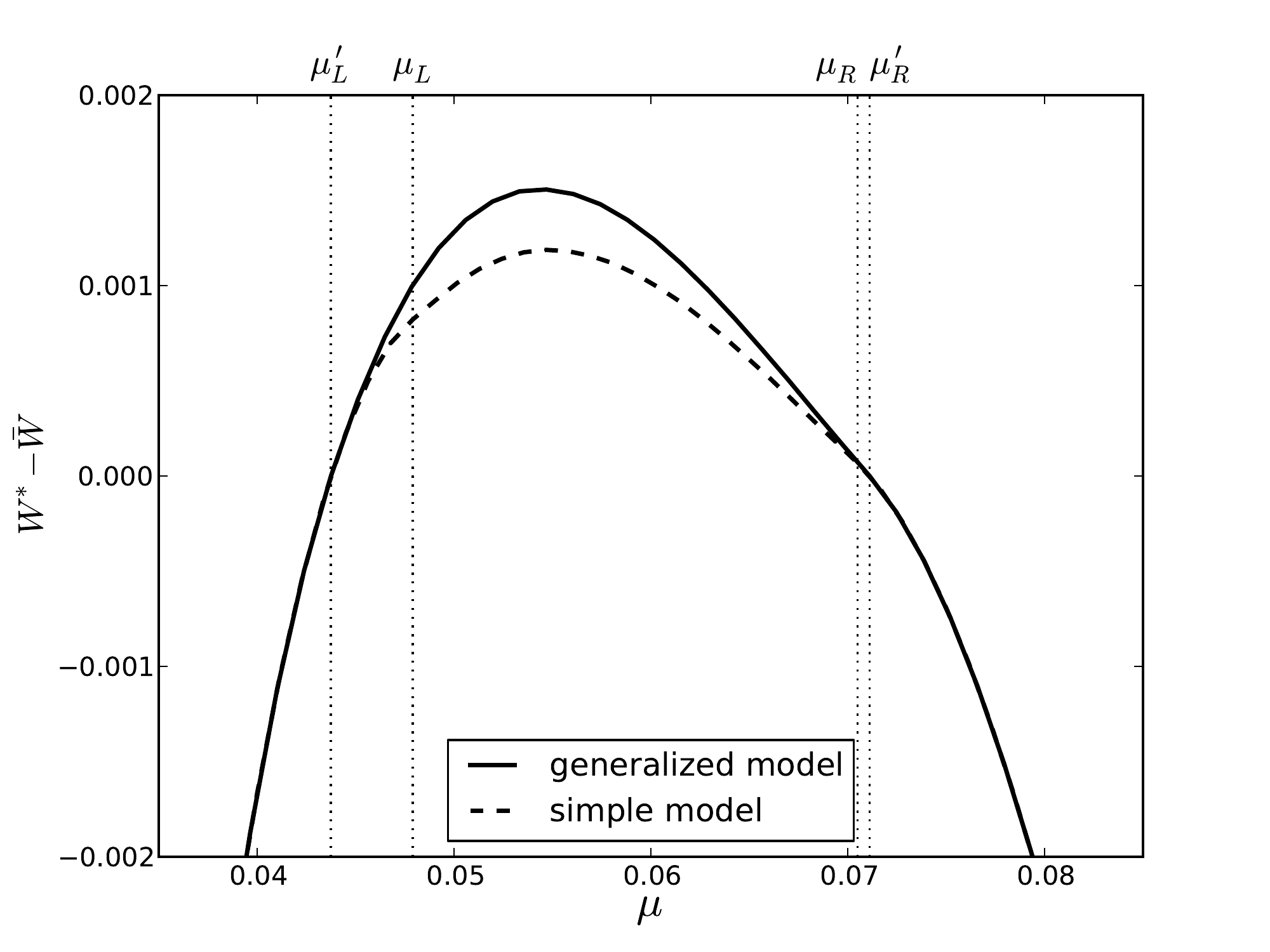}%
}
\caption{\small Generalized model with parent-dependent phenotype distribution. The birth and death rates of each phenotype are $\beta_\textrm{A} = 3$, $\delta_\textrm{A} = 2$, and $\beta_\textrm{B} = 1$, $\delta_\textrm{B} = 0.25$; the carrying capacity of each local patch is $K = 100$. (\textbf{a}) Optimal phenotypic switching probabilities $\sigma_a^*$ ($a = \textrm{A}, \textrm{B}$) with respect to the dispersal rate $\mu$. For comparison, dashed lines show the optimal phenotype distribution $\rho^*$ and $(1-\rho^*)$ given by the simple parent-independent model presented in the main text. (\textbf{b}) Maximum asymptotic expansion rate $W^*$ versus dispersal rate $\mu$ in both the simple (dashed curve) and the generalized (solid curve) models. For clear comparison, we subtracted a same term $\bar{W}(\mu) \equiv \frac{\mu_R^\prime - \mu}{\mu_R^\prime - \mu_L^\prime} W^*(\mu_L^\prime) + \frac{\mu - \mu_L^\prime}{\mu_R^\prime - \mu_L^\prime} W^*(\mu_R^\prime)$ from each curve.}
\label{fig:parent}
\end{figure}

Instead of Eq.~(\ref{eq:birth-mixed}), the birth process is now described by:
\begin{align}
\mathbb{I}^a + \emptyset &\xrightarrow{\mathmakebox[0.5in]{\pi_{ba} \, \beta_a /K}} \mathbb{I}^a + \mathbb{I}^b , \label{eq:birth-mixed-parent}
\end{align}
whereas the death and dispersal processes remain the same. Accordingly, for the patch dynamics, Eqs.~(\ref{eq:patch-birthA}, \ref{eq:patch-birthB}) are replaced by
\begin{align}
\textrm{P}_{n,l} &\xrightarrow{\mathmakebox[0.2in]{\eta_{nl}}} \textrm{P}_{n+1,l+1} \; , \label{eq:patch-birthA-parent} \\
\textrm{P}_{n,l} &\xrightarrow{\mathmakebox[0.2in]{\theta_{nl}}} \textrm{P}_{n+1,l} \; , \label{eq:patch-birthB-parent}
\end{align}
where the new rate constants are $\eta_{nl} = (1-\sigma_\textrm{A}) \beta_\textrm{A} l (1-n/K) + \sigma_\textrm{B} \beta_\textrm{B} (n-l)(1-n/K)$, and $\theta_{nl} = \sigma_\textrm{A} \beta_\textrm{A} l (1-n/K) + (1-\sigma_\textrm{B}) \beta_\textrm{B} (n-l)(1-n/K)$. The asymptotic expansion rate $W$ can be calculated in the same way as in Sec.~\ref{app:asymp-mixed}, by replacing $\rho \, \beta_{nl}$ with $\eta_{nl}$ and replacing $(1-\rho) \beta_{nl}$ with $\theta_{nl}$.

In this generalized model, we maximize $W$ with respect to both $\sigma_\textrm{A}$ and $\sigma_\textrm{B}$. Their optimal values, $\sigma_\textrm{A}^*$ and $\sigma_\textrm{B}^*$, are plotted as functions of the dispersal rate $\mu$ in Fig.~\subref*{fig:sigma_opt}. There is a lower bound, $\mu_L^\prime$, below which $\sigma_\textrm{B}^* = 0$ and $\sigma_\textrm{A}^*$ can take any values between $0$ and $1$. It means that, for such very low dispersal rates, offspring of parental phenotype $\textrm{B}$ never switch to phenotype $\textrm{A}$, and this ``pure $\textrm{B}$'' subpopulation contributes the maximum asymptotic expansion rate for the species; meanwhile, there may coexist a nonzero subpopulation of phenotype $\textrm{A}$ if $\sigma_\textrm{A} < 1$, yet their number would be subdominant and does not affect the asymptotic expansion rate. Similarly, for dispersal rates above an upper bound, $\mu_R^\prime$, the optimal phenotypic switching probabilities are $\sigma_\textrm{A}^* = 0$ and $\sigma_\textrm{B}^* =$ any value between $0$ and $1$. It means that, for sufficiently high dispersal rates, the optimal strategy for the species is to have a dominant ``pure $\textrm{A}$'' subpopulation that never switch to phenotype $\textrm{B}$.

Note that the basic parent-independent model presented in the main text corresponds to $\sigma_\textrm{A} = 1 - \rho$ and $\sigma_\textrm{B} = \rho$, satisfying $\sigma_\textrm{A} + \sigma_\textrm{B} = 1$. The generalized model relaxes the last constraint by letting $\sigma_\textrm{A}$ and $\sigma_\textrm{B}$ vary independently. It can be seen from Fig.~\subref*{fig:sigma_opt} that their optimal values do not satisfy the constraint. For comparison, the optimal value $\rho^*$ obtained for the simple model is also shown in the figure, which satisfies $0 < \rho^* < 1$ for $\mu$ between two bounds $\mu_L$ and $\mu_R$. Since the generalized model is less constrained, one expects the maximum asymptotic expansion rate $W^*$ to be no less than that in the simple model. This is indeed the case, as illustrated in Fig.~\subref*{fig:W_opt}. The two curves coincide at values of the dispersal rate below $\mu_L^\prime$ and above $\mu_R^\prime$, because in those cases only one phenotype, $\textrm{B}$ and $\textrm{A}$ respectively, completely dominates the population in both the simple and the generalized models.

\end{document}